\documentclass[a4paper,11pt]{article}
\pdfoutput=1 % if your are submitting a pdflatex (i.e. if you have
\usepackage{jheppub} % for details on the use of the package, please
\usepackage{slashed}
\usepackage{amsfonts}
\usepackage{amsmath}
\usepackage{amssymb}
\usepackage{bbm}
\usepackage[vcentermath]{youngtab}
\usepackage[T1]{fontenc} % if needed
\usepackage{silence}

\newcommand{\beq}{\begin{eqnarray}}
\newcommand{\eeq}{\end{eqnarray}}

\newcommand{\bmp}{\noindent\begin{minipage}{16cm}}
\newcommand{\emp}{\end{minipage}\vskip 7mm} % 7mm untightened

\newcommand{\SU}{\mbox{SU}}
\newcommand{\SO}{\mbox{SO}}
\newcommand{\SP}{\mbox{Sp}}
\newcommand{\UU}{\mbox{U}}

\definecolor{pumpkin}{rgb}{1.0, 0.4, 0.0}

\title{Gravitational waves from $ \SU(N)/\SP(N) $ composite Higgs models}

\author[a]{Mads T. Frandsen,}
\author[b]{Matti Heikinheimo,}
\author[c]{Martin Rosenlyst,}
\author[a]{Mattias E. Thing}
\author[b]{and Kimmo Tuominen}

\affiliation[a]{$ \text{CP}^3 $-Origins, University of Southern Denmark, Campusvej 55, DK-5230 Odense M, Denmark}
\affiliation[b]{Helsinki Institute of Physics and Department of Physics, University of Helsinki,
P.O.Box 64, FI-00014 University of Helsinki, Finland}
\affiliation[c]{Rudolf Peierls Centre for Theoretical Physics, University of Oxford, 1 Keble Road, Oxford OX1 3NP, United Kingdom\\}

\emailAdd{frandsen@cp3.sdu.dk}
\emailAdd{matti.heikinheimo@helsinki.fi}
\emailAdd{martin.jorgensen@physics.ox.ac.uk}
\emailAdd{thing@cp3.sdu.dk}
\emailAdd{kimmo.i.tuominen@helsinki.fi}

\abstract{We study possible strong first-order electroweak phase transitions in Composite Higgs models and we quantify the part of parameter space that can be probed with future gravitational Wave experiments. We focus on models where the Composite Higgs sector arises from underlying four-dimensional strongly interacting gauge theories with fermions, and where the Standard Model fermion masses are induced via linear mixing terms with composite fermions -- the
so-called fermion partial compositeness framework. We perform our analysis for the general class of Composite Higgs models arising from $ N $ Weyl fermions in a pseudo-real representation of the new strongly interacting gauge group that dynamically triggers the global chiral symmetry breaking pattern $ \SU(N)\rightarrow \SP(N) $. The minimal model has $ N=4 $ and for $ N>4 $ the models feature complex scalar dark matter candidates arising as pseudo-Nambu-Goldstone bosons. We find a large number of points in the models parameter space which yield strong first-order electroweak phase transitions and identify the most important operators characterizing the strength of the phase transition. Almost all of these points are testable with future GW detectors such as LISA, Taiji, Tianqin, BBO, DECIGO and Ultimate-DECIGO.
}
\preprint{HIP-2023-1/TH}
\begin{document} 
\maketitle
\flushbottom

\section{Introduction}

In the Standard Model (SM) of particle physics, the origin of the Higgs sector and the related naturalness and triviality problems are important open questions. These questions are addressed in the Composite Higgs (CH) framework~\cite{Kaplan:1983fs,Dugan:1984hq}, where the Higgs boson arises as a composite pseudo-Nambu-Goldstone boson (PNGB) from a spontaneously broken global symmetry.~\footnote{In general, composite Higgs models were first proposed in Refs.~\cite{Terazawa:1976xx,Terazawa:1979pj}, but in these models the Higgs boson is not identified as a light pNGB candidate. } 

Due to this, the Higgs candidate can be naturally light with respect to the compositeness scale while other composite resonances remain heavy. Moreover, the couplings of the Higgs candidate can be SM-like and therefore these models remain in good agreement with current observations in a large part of the parameter space of the models.

We will further assume that the relevant global symmetry of the Higgs sector arises as the chiral symmetries $G$ of fermions of an underlying confining four-dimensional gauge-fermion theory. The strong interactions dynamically break the global symmetries $G$ to $H$, while the interactions with the SM fields break the symmetries further, ensuring the correct electroweak symmetry breaking (EWSB) pattern. Finally, we assume that the masses for the SM-fermions are induced by a linear mixing of the SM-fermions with composite fermions of the new strongly interacting sector, the so-called fermion Partial Compositeness (PC) mechanism~\cite{Kaplan:1991dc}. 

A well-known challenge in CH models is violations of the flavor changing neutral current (FCNC) constraints~\cite{Cacciapaglia:2020kgq} due to the effective four-fermion operators generating the SM-fermion masses, especially in the case of the heavy top quark mass. With the PC mechanism, the linear mixing of the top quark with the composite fermionic operators can have sufficiently large anomalous dimensions, resulting in a suppression of these FCNCs. For this reason, we will use the PC mechanism throughout this paper.

Our main goal in this work is to explore gravitational wave (GW) signals from strong first-order electroweak phase transitions (SFOEWPT) driven by the strong interactions of the new strongly interacting fermions (hyper-fermions). In particular we aim to study constraints on the parameter space of the models in light of gravitational wave (GW) experiments planned for the future. For concreteness, we will focus on CH model realizations where the PNGBs correspond to broken generators of cosets $ G/H=\SU(N)/\SP(N) $ (e.g. the minimal $ \SU(4)/\SP(4) $ CH model proposed in Ref.~\cite{Galloway:2010bp}~\footnote{The smaller $ \SO(5)\rightarrow \SO(4) $ coset is difficult to realize in a simple manner from a four-dimensional gauge-fermion theory but arises more simply in the five-dimensional holographic approach~\cite{Agashe:2004rs}.}). We find that almost all  CH models with SFOEWPTs are testable via GW signals at future detectors such as LISA~\cite{LISA:2017pwj}, Taiji~\cite{Hu:2017mde}, Tianqin~\cite{TianQin:2015yph}, BBO~\cite{Crowder:2005nr}, DECIGO~\cite{Kawamura:2006up,Kawamura:2011zz} and Ultimate-DECIGO (U-DECIGO)~\cite{Kawamura:2011zz}. 

Furthermore, in the CH framework with extended global symmetries ($N>4$), dark matter (DM) candidates arise as a composite PNGBs with relic densities produced either thermally~\cite{Ma:2015gra,Wu:2017iji,Cai:2018tet,Alanne:2018xli,Cacciapaglia:2019ixa} or non-thermally~\cite{Cai:2019cow,Cai:2020bhd,Cacciapaglia:2021aex,Rosenlyst:2021elz}.~\footnote{In the CH framework with $ N=4 $, the only possible dark matter candidate of the five composite PNGBs is the pseudo-scalar $ \eta $, which may be protected by the parity transformation $ \eta\rightarrow -\eta $ and therefore will be stable. However, this apparent discrete symmetry is not a symmetry of the underlying theory, since it breaks at the effective Lagrangian level via topological-induced terms. } Moreover, with such global symmetries, an additional composite pNGB Higgs doublet can be employed to loop-induce neutrino masses using a scotogenic mechanism as demonstrated in Ref.~\cite{Cacciapaglia:2020psm}. We show how some of these CH and dark matter models can be tested via the GW results we consider in this paper.

In the effective Lagrangian description of the models we identify an effective Higgs potential term generated by the PC mechanism which is particularly important for GW signals. 
The term originates from a mixture of four-fermion operators containing the left- and right-handed top quarks and is proportional to a coefficient $ C_{LR} $ that must be determined non-perturbatively, e.g. via lattice simulations. Finally, we show that when $ C_{LR}>0 $ the regions of the parameter space satisfying the SFOEWPT condition increase significantly. 
 
To compute the relevant parameters that determine the SFOEWPT and to compute the GW signals, we use the Python package $\textsf{CosmoTransitions} $~\cite{Wainwright:2011kj} with custom modifications to extract the needed parameters and select the relevant phase of the transition.

\section{SU(N)/Sp(N) Composite Higgs Models}
\label{sec: SU(2N)/Sp(2N) composite Higgs models}

We begin by reviewing four-dimensional hyper-fermion sectors charged under a new strongly
interacting gauge group $G_{\rm HC}$. The complete gauge group of the models considered is then \beq
G_{\rm HC} \otimes \text{SU}(3)_{\rm C}  \otimes \text{SU}(2)_{L}  \otimes \text{U}(1)_{Y}\,.
\eeq 

For models with a single fermionic representation of $ G_{\rm HC}$, the symmetry breaking patterns are known~\cite{Witten:1983tx,Kosower:1984aw}: Given $ N $ Weyl fermions transforming as the $\mathcal{R}_{\rm HC}$ representation of $ G_{\rm HC} $, the three possible classes of vacuum cosets $G/H$ are $ \SU(N)/\SP(N) $ for pseudo-real $\mathcal{R}_{\rm HC}$, $ \SU(N)\otimes \SU(N)\otimes \UU(1)/ \SU(N)\otimes \UU(1)$ for complex $\mathcal{R}_{\rm HC}$~\cite{Peskin:1980gc} and $ \SU(N)/\SO(N) $ for real $\mathcal{R}_{\rm HC}$.
The minimal CH cosets that can provide a SM-like composite PNGB Higgs multiplet with custodial symmetry, within these three classes have $N=4$ in both the pseudo-real~\cite{Galloway:2010bp} and the complex cases~\cite{Ma:2015gra}, and $N=5$ in the real case~\cite{Dugan:1984hq}. 

The pseudo-real coset $\SU(4)/\SP(4)$ is the most minimal as its PNGB spectrum includes only five composite PNGB states: a Higgs doublet $H$ and a pseudo-scalar singlet $\eta$. At the effective Lagrangian level, $ \eta $ is protected by approximate parity symmetry ($ \eta\rightarrow -\eta $), which is broken only by induced topological terms in the effective Lagrangian~\cite{Bauer:2017ris}. The prospects of $ \eta $ as a composite DM candidate have been studied in e.g. Ref.~\cite{Frigerio:2012uc}. In models with $N > 4$, the extended PNGB coset $\SU(N)/\SP(N)$ provides complex scalar composite PNGB DM candidates protected by unbroken $\UU(1)$ symmetries analogous to the baryon number in QCD. The $N = 6$ case is studied in Ref.~\cite{Cai:2018tet,Cai:2019cow,Cai:2020bhd} and the $N = 8$ in Ref.~\cite{Cacciapaglia:2021aex}. 
While relevant for both of these model examples, the results we obtain in this paper have wider applicability: we consider CH description which includes the general class of CH models with cosets 
$\SU(N)/\SP(N)$, where the vacuum is aligned in the direction of the PNGB Higgs candidate. 

We note that alternative model building can be pursued with other cosets such $ \SU(4)\otimes \SU(4)\otimes \UU(1)/\SU(4) \otimes \UU(1)$~\cite{Ma:2015gra,Wu:2017iji} and $ \SU(6)/\SO(6)$~\cite{Cacciapaglia:2019ixa}, but we do not focus on these possibilities in this work.

\subsection{Underlying model Lagrangian}
\label{sec: The fermion content}

In this section, we introduce the new strongly interacting hyper-fermions. We consider $N=2N_L+2N_R$ Weyl hyper-fermions with $ \SU(2)_{ L}\otimes \UU(1)_{ Y} $ charges
\beq
\label{eq:hyperew}
 &&\Psi^i_{\alpha,a}=\begin{pmatrix}\psi^{2i-1}_{\alpha,a}\\ \psi^{2i}_{\alpha,a}\end{pmatrix}\sim(2,0)\,, \quad \Upsilon^j_{\alpha,a}=(\chi^{2j-1}_{\alpha,a},\chi^{2j}_{\alpha,a})\sim(1,\mp 1/2)\,,
\eeq 
where $\alpha=1,2$ is the left-handed spinor index, $a$ is the representation index for 
$G_{\rm HC}$
and $ i=1,\dots, N_L $ and $ j=1,\dots, N_R $. The spinor $\Psi^i$ transforms as a doublet under the $i$'th $\SU(2)_{L }$ global symmetry in $\SU(N)$ and the spinor $\Upsilon^j$ transforms as a doublet under the $j$'th $\SU(2)_{R }$ global symmetry in $\SU(N)$ .
The case $N_L=N_R=1$ results in the minimal $ \SU(4)/\SP(4) $ Composite Higgs (CH) model~\cite{Galloway:2010bp}. We assume that the hyper-fermions $Q$ transform under the fundamental representation of $ G_{\rm HC}=\SP(2N_{\rm HC}) $ or spin representation of $ G_{\rm HC} =\SO(N_{\rm HC})$ which is the pseudoreal representation, resulting in the chiral symmetry breaking pattern $\SU(N)\rightarrow \SP(N) $ when the fermions confine. These gauge groups $G_{\rm{HC}}$ are determined by the full global symmetry 
breakig pattern which we will soon discuss.  

For later convenience, we will arrange the $ N $ Weyl spinors in a  vector $Q$ transforming under the fundamental representations of the  $ \SU(N)$ chiral symmetry group:  \beq
 Q_{\alpha, a}^I\equiv \begin{pmatrix}Q_{4}\\ Q_L\\ Q_R 
  \end{pmatrix}_{\alpha, a}
 \,, \label{eq: SU(2N) vector}
\eeq where the $I=1, ..., N$ ``flavor'' index runs as defined by Eq.~\eqref{eq: SU(2N) vector} together with the definitions
\begin{equation}
\begin{aligned}  
 Q_{4}& = (\psi^1,\psi^2,\chi^1,\chi^2)^T \,, \\
  Q_{L} &= (\psi^3,\psi^4,\dots, \psi^{2N_L-1},\psi^{2N_L})^T\,, \quad N_L>1 \\
   Q_{R} &= (\chi^3,\chi^4,\dots,\chi^{2N_R-1},\chi^{2N_R})^T , \quad N_R >1
 \,.
\end{aligned}
\end{equation} 

In general, we may also introduce $ N_\Lambda $ SM-singlet hyper-fermions $ \lambda^k $ with $k=1,\dots,N_\Lambda$~\cite{Ryttov:2008xe,Alanne:2018xli,Rosenlyst:2021elz}. However, such extra SM-singlet hyper-fermions have no influence on the generation of GWs we will study in this work and therefore we do not consider them further. The origin of fermion masses requires extending the CH sector. To address this using the PC mechanism, we introduce a new species of fermions $\chi_t$, transforming under the two-index antisymmetric representation of $ G_{\rm HC}=\SP(2N_{\rm HC}) $ or under the fundamental representation under $ G_{\rm HC} =\SO(N_{\rm HC})$.

For the top quark, it is enough to introduce a $\SU(2)_L$ vector-like 
$\chi_t$-pair with hypercharge $+2/3$ transforming as a fundamental of the ordinary $ \SU(3)_C $ color gauge group (QCD). Models of this types were first proposed in Refs.~\cite{Barnard:2013zea,Ferretti:2013kya} and our model is an extension of the one in Ref.~\cite{Barnard:2013zea}. 

Now we can write the underlying Lagrangian for the $ N $ Weyl hyper-fermions with EW charges as detailed in Eq.~\eqref{eq:hyperew}, and arranged into a $ \SU(N) $ vector, $Q$, shown in Eq.~(\ref{eq: SU(2N) vector}). 
 As already discussed in the previous section, we set $N_\Lambda=0$. In terms of this multiplet, the underlying gauge-fermion Lagrangian of the CH model can be written as \beq 
\mathcal{L}_{\rm CH}= Q^\dagger i \gamma^\mu D_\mu Q - \frac{1}{2} \left(Q^T M_Q Q +{\rm h.c.} \right)
	 + \mathcal{L}_{\rm PC}\,, 
\label{eq: Basic Lagrangian (UV)}
\eeq where the covariant derivative includes the $G_{\rm HC}$, $\SU(2)_{L}$ and $ \UU(1)_{Y} $ gauge bosons. The mass term is $2\times 2$ block diagonal of the form $M_Q=\text{Diag}(\overline{m}_1i\sigma_2,-\overline{m}_2i\sigma_2,\overline{m}_3i\sigma_2,\dots)$ 
and consists of $N_{L}$ independent mass matrices $ \overline{m}_{i} i\sigma_2 $ for each $ \SU(2)_{L} $ hyper-fermion doublet and $N_{R}$ independent mass matrices $ -\overline{m}_{j} i\sigma_2 $ for each $ \SU(2)_{R} $ hyper-fermion doublet,
respectively, arranged as in Eq.~(\ref{eq: SU(2N) vector}) where $ \sigma_2 $ is the second Pauli matrix. It contributes to the correct vacuum alignment, parameterized by an angle $\theta$, between an EW unbroken vacuum and a Technicolor type vacuum. 

We  arrange the new QCD colored hyper-fermions $\chi_t$ into a $ \SU(6)_\chi $ vector, spontaneously breaking to $ \SO(6)_\chi $ upon the condensation. Thus, we end up with CH models with the coset structure 
\beq \label{eq: the chiral symmetry breaking}
\frac{\SU(N)}{\SP(N)}\otimes\frac{\SU(6)_\chi}{\SO(6)_\chi}\otimes \frac{\UU(1)_Q \otimes \UU(1)_\chi}{\varnothing}\,,
\eeq
where one of the subgroups of $ \UU(1)_Q \otimes \UU(1)_\chi $ has an anomaly with the hypercolor symmetry group, $ G_{\rm HC} $, while the other one is an anomaly-free subgroup $ \UU(1)_\sigma $. Moreover, the QCD gauge group $ \SU(3)_C $ will be identified as a gauged subgroup of the unbroken group $ \SO(6)_\chi $. The global symmetry breaking pattern in Eq.~(\ref{eq: the chiral symmetry breaking}) determines the possible hypercolor groups~\cite{Ferretti:2013kya}, to be $ G_{\rm HC}=\SP(2N_{\rm HC}) $ with $ 2\leq N_{\rm HC} \leq 18 $ or $ G_{\rm HC}=\SO(N_{\rm HC})$ with $ N_{\rm HC}=11,13 $.
Therefore, the minimal choice is $ G_{\rm HC}=\SP(4)_{\rm HC} $ for $ N \lesssim  66 $ ~\cite{Sannino:2010ia}.  

The four-fermion interactions that will generate the PC operators we consider are~\cite{Alanne:2018wtp}: 
\beq \label{eq: top PC-operators}
{\mathcal L}_{\rm PC}\supset
 \frac{\widetilde{y}_{L}}{\Lambda_t^2} q^{\alpha\dagger}_{L,3} (Q^\dagger P_q^\alpha Q^*\chi_{t}^\dagger) + \frac{\widetilde{y}_{R}}{\Lambda_t^2}t_{R}^{c\dagger} (Q^\dagger P_t Q^* \chi_{t}^\dagger)+\rm h.c.\,, 
\eeq  
where $ q_{L,3} $ and $ t_R $ are the third generation of the quark doublets and the top singlet, respectively, and $P_q$ and $P_t$ are spurions that project onto the appropriate  components in the multiplet $Q$ in Eq.~(\ref{eq: SU(2N) vector}). For both the left- and right-handed top quark, we choose the spurions to transform as the two-index antisymmetric of the unbroken chiral symmetry subgroup $ \SP(N) $ as the minimal possible choice. Concretely, the left-handed spurions are given by the matrices
\begin{equation}
\begin{aligned}  
P_{q,ij}^1=\frac{1}{\sqrt{2}}(\delta_{i1}\delta_{j3} - \delta_{i3}\delta_{j1})\,, \quad \quad P_{q,ij}^2=\frac{1}{\sqrt{2}}(\delta_{i2}\delta_{j3} - \delta_{i3}\delta_{j2})\,, \end{aligned}  
\end{equation} 
while the right-handed spurion can be written as \begin{equation}
\begin{aligned} \label{eq: Pt matrix} P_t = -\frac{1}{\sqrt{2(N_L+N_R)}}\text{Diag}(P_{t4},P_{tL},P_{tR})\,. \end{aligned} 
\end{equation} Here $P_{t4}\equiv \text{Diag}(i\sigma_2,-i\sigma_2) $ representing the $\SU(4)/\SP(4)$ subset of $ \SU(N)/\SP(N) $, while $P_{tL,tR}\equiv \pm \text{Diag}(i\sigma_2,\dots,i\sigma_2) $ are block diagonal matrices containing $N_{L,R}-1$ of the $\pm i\sigma_2$ matrix on the diagonal, respectively.

\subsection{The effective Lagrangian of the PNGBs}

At the condensation scale $ \Lambda_{\rm HC}\sim 4\pi f $, where $f$ is the decay constant of the composite PNGBs, the hyper-fermions can form an antisymmetric and SM gauge invariant condensate of the form
\beq
 \langle Q^I_{\alpha,a}Q^J_{\beta,b}\rangle\epsilon^{\alpha\beta}\epsilon^{ab}\sim \Sigma_{Q0}^{IJ}=\text{Diag}(P_{t4},P_{tL},P_{tR}) \,, \label{eq: EW unbroken vacuum}
\eeq
where the diagonal matrices $P_{t4}$, $P_{tL}$ and $P_{tR}$ are given below Eq.~(\ref{eq: Pt matrix}).  

This condensate breaks $ \SU(N)\rightarrow \SP(N) $ due to the antisymmetry of $\Sigma_{Q0}^{IJ}$, resulting in a set of PNGBs $ \pi_A $ corresponding to the broken generators, $ X_A $, with $ A=1,\dots,N^2/2-N/2-1 $. 

The five broken generators in the subset $ \SU(4)/\SP(4) $ of $ \SU(N)/\SP(N) $ are given by	\beq
		&&X_{1}=\frac{1}{2\sqrt{2}}\begin{pmatrix}0&\sigma^{3}&0 & \cdots\\ \sigma^{3}&0&0& \cdots\\0&0&0& \cdots \\ \vdots & \vdots & \vdots & \ddots \end{pmatrix}\,,\quad \quad \quad X_2=\frac{1}{2\sqrt{2}}\begin{pmatrix}0&i1_{2\times2}&0&\cdots\\-i1_{2\times2}&0&0&\cdots\\0&0&0&\cdots \\\vdots & \vdots & \vdots & \ddots \end{pmatrix}\,, \nonumber \\ &&X_{3}=\frac{1}{2\sqrt{2}}\begin{pmatrix}0&\sigma^{1}&0 & \cdots\\ \sigma^{1}&0&0& \cdots\\0&0&0& \cdots \\ \vdots & \vdots & \vdots & \ddots \end{pmatrix}\,,\quad \quad \quad X_{4}=\frac{1}{2\sqrt{2}}\begin{pmatrix}0&\sigma^{2}&0 & \cdots\\ \sigma^{2}&0&0& \cdots\\0&0&0& \cdots \\ \vdots & \vdots & \vdots & \ddots \end{pmatrix}, \\ \nonumber && X_5=\frac{1}{2\sqrt{2}}\begin{pmatrix}1_{2\times2}&0&0 & \cdots\\0&-1_{2\times2}&0& \cdots\\0&0&0& \cdots \\ \vdots & \vdots & \vdots & \ddots\end{pmatrix}\,, \nonumber
\eeq where the PNGB Higgs candidate $h$ and the pseudo-scalar $\eta $ are given by \begin{equation}
\begin{aligned}
\frac{h}{f}\equiv\frac{\pi_4}{\sqrt{\pi_4^2+\pi_5^2}}\sin\left(\frac{\sqrt{\pi_4^2+\pi_5^2}}{f}\right)\,, \quad\quad  \frac{\eta}{f}\equiv\frac{\pi_5}{\sqrt{\pi_4^2+\pi_5^2}}\sin\left(\frac{\sqrt{\pi_4^2+\pi_5^2}}{f}\right)\,,
\end{aligned}
\end{equation} respectively, while $\pi_{1,2,3}$ associated with $X_{1,2,3}$ are the Goldstones eaten by the $W^\pm$ and $Z$ gauge bosons when the Higgs boson achieves a VEV via vacuum misalignment. Furthermore, the unbroken generators $ T_{\overline{A}}=\{ T_L^a, T_R^a, T_r^i\} $ with $ \overline{A}=1,\dots,N(N+1)/2 $, $ a=1,2,3 $ and $ i=1,\dots,N^2/2+N/2-6 $, in which $ T_{L,R}^a $ belong to the custodial symmetry subgroup $ \SO(4)_C\cong \SU(2)_L \otimes \SU(2)_R $ while $ T_R^i $ belong to the coset $ \SP(N)/\SO(4)_C $. We identify $ T_L^a $ and $ T_R^3 $ as the generators for the $ \SU(2)_L $ and $ \UU(1)_Y $ gauge groups, respectively, where their explicit matrix form are given by \begin{equation}
\begin{aligned}
&T_L^a=\text{Diag}\left(\sigma^a,0,\sigma^a,\dots,\sigma^a,0,\dots,0\right)\,, \\ 
&T_R^3 = \text{Diag}\left(0,-(\sigma^{3})^T,0,\dots,0,-(\sigma^{3})^T,\dots,-(\sigma^{3})^T \right) \label{eq: SU(2)L and U(1)Y matrices}
\end{aligned}
\end{equation} For the minimal coset $ \SU(4)/\SP(4) $, the explicit matrix forms of all these generators are listed in Ref.~\cite{Galloway:2010bp}. 

We parameterize the PNGBs via the exponential map $ U_Q=\exp[i\pi_A X_A /f] $, which in unitary gauge ($\pi_{1,2,3}\rightarrow 0$) can be written as~\cite{Dong:2020eqy}
\begin{equation}
\begin{aligned}
 U_Q=\left(\begin{array}{ccc} (c_Q+i\frac{\pi_5}{\pi_Q}s_Q)\mathbf{1}_2 & i\sigma_2\frac{\pi_4}{\pi_Q}s_Q & \cdots \\ i\sigma_2\frac{\pi_4}{\pi_Q}s_Q & (c_Q-i\frac{\pi_5}{\pi_Q}s_Q)\mathbf{1}_2 & \cdots\\ \vdots & \vdots & \ddots \end{array}\right)\,, \label{eq: UQ matrix}
\end{aligned}
\end{equation} where $ c_Q \equiv \cos (\pi_Q/(2f)) $ and $ s_Q \equiv \sin (\pi_Q/(2f)) $ with $ \pi_Q\equiv \sqrt{\pi_4^2+\pi_5^2} $ and only the SU(4)/Sp(4) states have been explicitly included. Finally, the explicit form of the PNGB matrix with coset $ \SU(N)/\text{Sp}(N) $ is given by $ \Sigma_Q=U_Q^2 \Sigma_{Q0} $. 

Below the compositeness scale $ \Lambda_{\rm HC} $, Eq.~\eqref{eq: Basic Lagrangian (UV)} is replaced by the effective Lagrangian:
\beq
    \label{eq: effLag}
    \mathcal{L}_\mathrm{eff}=\mathcal{L}_{\mathrm{kin}}-V_{\mathrm{eff}}\,.
\eeq 
Here $ \mathcal{L}_{\mathrm{kin}} $ is the usual leading order ($\mathcal{O} (p^2)$) chiral Lagrangian~\cite{Cacciapaglia:2020kgq},\beq
    \label{eq:effLag kin}
\mathcal{L}_{\mathrm{kin},p^2}=\frac{f^2}{4}\text{Tr}[d_\mu d^\mu]\,,
\eeq where the $ d $ and $ e $ symbols of this model are given by the Maurer-Cartan form \begin{equation}
\begin{aligned}
iU_Q^\dagger D_\mu U_Q \equiv d_\mu^A X^A + e_\mu^{\overline{A}} T^{\overline{A}}\equiv d_\mu + e_\mu\,, \label{eq: d and e symbols}
\end{aligned}
\end{equation} with the gauge covariant derivative \begin{equation}
\begin{aligned}
D_\mu = \partial_\mu - i g W_\mu^a T_L^a - i g^\prime B_\mu T_R^3\,.
\end{aligned}
\end{equation} The kinetic Lagrangian term in Eq.~(\ref{eq:effLag kin}) provides kinetic terms and self-interactions for the PNGBs. In addition, it will induce masses for the EW gauge bosons and their couplings with the PNGBs (including the SM Higgs identified as $ h $), \begin{equation}
\begin{aligned} &m_W^2 = \frac{1}{4}g^2 f^2 \sin^2\theta\,, \quad\quad m_Z^2=m_W^2/c^2_{\theta_W}\,, \\
&g_{hWW}=\frac{1}{4} g^2 f\sin  2\theta=g_{  hWW}^{\rm SM}\cos \theta \,,\quad\quad g_{ hZZ}=g_{ hWW}/c^2_{\theta_W}\,, \label{eq: WZ masses and SM VEV}
\end{aligned}
\end{equation} where $ \theta_W $ is the Weinberg angle, $ g $ is the weak $ \SU(2)_{ L} $ gauge coupling and \begin{equation}
\begin{aligned} 
\sin\theta \equiv \frac{\langle h\rangle}{f}  \label{SM VEV}
\end{aligned}
\end{equation} with $\langle h\rangle= v_{\rm SM}=246~\text{GeV}$ after the EWSB. The vacuum misalignment angle $ \theta $ parametrizes the deviations of the CH Higgs couplings to the EW gauge bosons with respect to the SM Higgs. These deviations are constrained by direct LHC measurements~\cite{deBlas:2018tjm} of this coupling which imply an upper bound of $ s_\theta \lesssim 0.3 $. EW precision measurements also impose an upper limit which has been found to be stricter $ s_\theta \lesssim 0.2 $~\cite{Cacciapaglia:2020kgq}. However, in Ref.~\cite{Frandsen:2022xsz}, it is shown that the constraints on $ s_\theta $ from EW precision measurements are alleviated when all contributions from the composite fermion resonances in the CH models are included. Furthermore, there exists a lower bound on the lightest top-partner (the singlet $T_1 $ defined in Eq.~(\ref{eq: explicit embeddings top partners})) with $ m_{T_1}\gtrsim 1.3 $~TeV, set by CMS and ATLAS~\cite{ATLAS:2018ziw,CMS:2018zkf,CMS:2017ynm,CMS:2019eqb}. Referring to Figure~3 in Ref.~\cite{Frandsen:2022xsz}, this bound does not impose further restrictions on the parameter space explored in this paper.

The generation of the gravitational waves is controlled by the effective potential $V_{\mathrm{eff}}$, which receives contributions from the EW gauge interactions (the spin-1 resonances), the SM fermion couplings to the strong sector (the fermion resonances) and the vector-like masses of the hyper-fermions. At leading order, each source of symmetry breaking contributes independently to the effective potential:
\beq
V_{\text{eff}}&=& V_{\text{gauge}}+V_{\text{top}}+V_\text{m}+\dotsc\,, \label{eq: eff pot total}
\eeq 
where the dots are left to indicate the presence of mixed terms at higher orders, or the effect of additional UV operators. The potential contributions can be further split into IR and UV terms from the EW gauge interactions (given in Eqs.~(\ref{eq: pot from EW gauge int}) and~(\ref{eq: pot from EW gauge int UV})) and the top quark couplings to the strong sector (in Eqs.~(\ref{eq: pot from top int}) and~(\ref{eq: pot from top int UV})) as follows

\begin{equation}
\begin{aligned}  
&V_{\rm gauge}=  V_{\rm gauge}^{\rm IR}+V_{\rm gauge}^{\rm UV}\,, \\ & V_{\rm top}=  V_{\rm top}^{\rm IR}+V_{\rm top}^{\rm UV}\,. 
\end{aligned}
\end{equation} Finally, the potential contribution $ V_{\rm m} $ from the vector-like masses is given by Eq.~(\ref{eq: vector-like masses pot}). 

In the following two sections, we will consider the effective theory of the composite spin-1 and fermion resonances, resulting in the potential contributions $ V_{\rm gauge} $ and $ V_{\rm top} $, respectively, discussed in Section~\ref{sec: The scalar potential and the condition of SFOEWPT}.

%%%%%%%%%%%%%%%%%%%%%%%%%%%%%%%%%%%%%%%%%%%%%%%%%%%%%%%%%%%%%%%%%%%%%%%%%%%
\subsection{The effective Lagrangian of the spin-1 resonances}
\label{sec: The effective theory of the spin-1 resonances}
%%%%%%%%%%%%%%%%%%%%%%%%%%%%%%%%%%%%%%%%%%%%%%%%%%%%%%%%%%%%%%%%%%%%%%%%%%%

The composite spin-1 resonances in these CH models, analogous to the $ \rho $- and $a_1$-mesons in the QCD, can be organized into an adjoint multiplet $ V_\mu =V_\mu^{\overline{A}} T^{\overline{A}} $ of the unbroken $ \SP(N) $ with $ \overline{A}=1,\dots,N(N+1)/2 $ and another multiplet $ A_\mu = A_\mu^A X^A $ in $ \SU(N)/\SP(N) $ with $ A=1,\dots,N(N-1)/2-1 $ (i.e. the two-index antisymmetric representation of $ \SP(N) $). Here $ T^{\overline{A}} $ and $ X^A $ are, respectively, the unbroken generators of $ \SP(N) $ and broken generators in $ \SU(N)/\SP(N) $. For the minimal CH model with the coset $ \SU(4)/\SP(4) $, these objects decompose into \begin{equation}
\begin{aligned}
&\left(\begin{array}{l}\textbf{10} \rightarrow \textbf{3}_{0}\oplus \textbf{1}_{1}\oplus \textbf{1}_{0}\oplus \textbf{1}_{-1}\oplus \textbf{2}_{1/2}\oplus \textbf{2}_{-1/2}\\ V \rightarrow V_L \oplus V_R^+ \oplus V_R^0 \oplus V_R^- \oplus V_D \oplus \widetilde{V}_D  \end{array}\right)\,, \quad \left(\begin{array}{l}\textbf{5} \rightarrow \textbf{2}_{1/2}\oplus \textbf{2}_{-1/2}\oplus \textbf{1}_{0}\\ A \rightarrow A_D \oplus \widetilde{A}_D \oplus A_S  \end{array}\right)
\end{aligned}
\label{Eq:vecs}
\end{equation} under the decomposition $ \SP(4)\rightarrow \SU(2)_L \otimes \UU(1)_Y $, where $ \widetilde{V}_D = i\sigma_2 V_D^* $ and $ \widetilde{A}_D=i\sigma_2 A_D^* $. These resonances are embedded in the multiplets $ V_\mu $ and $ A_\mu $ of CH models with non-minimal cosets $ \SU(N)/\SP(N) $. 

In the following, we will use the hidden gauge symmetry formalism to describe the spin-1 resonances, as done in Ref.~\cite{BuarqueFranzosi:2016ooy} for the minimal CH model. We, therefore, formally extend the global symmetry group $ \SU(N) $ to $ \SU(N)_1\otimes \SU(N)_2 $. We embed the elementary SM gauge bosons in $ \SU(N)_1 $ and the heavy composite spin-1 resonances in $ \SU(N)_2 $. In the effective Lagrangian, the $ \SU(N)_i $ ($ i=1,2 $) symmetries are spontaneously broken to $ \SP(N)_i $ via the introduction of two nonlinear representations $ U_i $ containing $ N(N-1)/2-1 $ PNGBs each (similar to the PNGB matrix $ U_Q $ in Eq.~(\ref{eq: UQ matrix})), which are given by \begin{equation}
\begin{aligned}
U_1=\exp\left(\frac{i}{f_1}\sum_{A=1}^{N(N-1)/2-1} \pi_1^A X^A\right)\,, \quad \quad U_2=\exp\left(\frac{i}{f_2}\sum_{A=1}^{N(N-1)/2-1} \pi_2^A X^A\right)\,.
\end{aligned}
\end{equation} They transform as $ U_i \rightarrow g_i U_i h(g_i,\pi_i)^\dagger $, where $ g_i $ and $ h(g_i,\pi_i) $ are group elements of $ \SU(N)_i $ and $ \SP(N)_i $, respectively.  Furthermore, the breaking of $ \SP(N)_1\otimes \SP(N)_2 $ into the chiral subgroup $ \SP(N) $ is described by the field \begin{equation}
\begin{aligned}
K=\exp\left(\frac{i}{f_K}\sum_{\overline{A}=1}^{N(N+1)/2}k^{\overline{A}} T^{\overline{A}}\right)\,,
\label{eq: K fields}
\end{aligned}
\end{equation} containing $ N(N+1)/2 $ PNGBs corresponding to the adjoint generators of $ \SP(2N) $ and transforming like $ K\rightarrow h(g_1,\pi_1) K h(g_2,\pi_2)^\dagger $. 

To write down the effective Lagrangian for the spin-1 resonances and $ K $, we define the gauged Maurer-Cartan one-forms as \begin{equation}
\begin{aligned}
\Omega_{i,\mu}&\equiv iU_i^\dagger D_\mu U_i \equiv d_{i,\mu}^A X^A + e_{i,\mu}^{\overline{A}} T^{\overline{A}}\equiv d_{i,\mu} + e_{i,\mu} \,, 
\label{eq: d and e symbols vec res}
\end{aligned}
\end{equation} where the covariant derivatives are given by 
\begin{equation}
\begin{aligned}
&D_\mu U_1 = \left(\partial_\mu -i g W_\mu^a T_L^a - i g^{\prime} B_\mu T_R^3\right)U_1\,, \\
& D_\mu U_2 = \left(\partial_\mu -i \widetilde{g} V_\mu^{\overline{A}}T^{\overline{A}} - i \widetilde{g} A_\mu^A X^A \right)U_2\,,
\end{aligned}
\end{equation} such that the heavy spin-1 vectors are formally introduced like gauge fields similar to the SM gauge fields. The covariant derivative of $ K $ has the form \begin{equation}
\begin{aligned}
D_\mu K =\partial_\mu K - i e_{1,\mu} K + iK e_{2,\mu}\,.
\end{aligned}
\end{equation} Here the $ d $ and $ e $ symbols transform under $ \SU(N)_i $ as follows \begin{equation}
\begin{aligned}
&d_{i,\mu} \rightarrow h(g_i,\pi_i) d_{i,\mu} h(g_i,\pi_i)^\dagger\,,  
\quad \quad e_{i,\mu} \rightarrow h(g_i,\pi_i) (e_{i,\mu}+i\partial_\mu)h(g_i,\pi_i)^\dagger\,.
\end{aligned}
\end{equation} 

Thus, the effective Lagrangian of the spin-1 resonances is given by \begin{equation}
\begin{aligned}
&\mathcal{L}_{\rm gauge} = -\frac{1}{2g^2} \text{Tr}[W_{\mu\nu}W^{\mu\nu}]-\frac{1}{2g^{\prime 2}}\text{Tr}[B_{\mu\nu}B^{\mu\nu}] - \frac{1}{2\widetilde{g}} \text{Tr}[V_{\mu\nu}V^{\mu\nu}] - \frac{1}{2\widetilde{g}} \text{Tr}[A_{\mu\nu}A^{\mu\nu}] \\ &\quad+ \frac{1}{2} f_1^2 \text{Tr}[d_{1,\mu}d^{\mu}_1]+ \frac{1}{2} f_{2}^2 \text{Tr}[d_{2,\mu}d^{\mu}_{2}]+r f_{2}^2 \text{Tr}[d_{1,\mu} K d^{\mu}_{2}K^\dagger]+ \frac{1}{2} f_K^2 \text{Tr}[D_\mu K D^\mu K^\dagger]\,,
\label{eq: VA lag}
\end{aligned}
\end{equation} where  \begin{equation}
\begin{aligned}
&W_{\mu\nu}=\partial_\mu W_\nu - \partial_\nu W_\mu -i g[W_\mu, W_\nu]\,,\quad \quad 
B_{\mu\nu}=\partial_\mu B_\nu - \partial_\nu B_\mu\,, \\ 
&V_{\mu\nu}=\partial_\mu V_\nu - \partial_\nu V_\mu -i \widetilde{g}[V_\mu, V_\nu]\,,\phantom{...} \quad \quad \quad 
A_{\mu\nu}=\partial_\mu A_\nu - \partial_\nu A_\mu -i \widetilde{g}[A_\mu, A_\nu]
\end{aligned}
\end{equation} with $ W_\mu = W_\mu^a T^a_L $. We have omitted terms containing a scalar singlet $ \sigma $ resonance, included in Ref.~\cite{BuarqueFranzosi:2016ooy}, as we here assume this resonance is heavy. 

In terms of the above Lagrangian parameters, we define the mass parameters \begin{equation}
\begin{aligned}
m_V\equiv \frac{\widetilde{g}f_K}{\sqrt{2}}, \quad \quad m_A\equiv \frac{\widetilde{g}f_1}{\sqrt{2}}\,.  
\label{eq: mV and mA mass parameters}
\end{aligned}
\end{equation} For the minimal CH model with coset $ \SU(4)/\SP(4) $, the masses of $ V_D $ and $ \widetilde{V}_D $ are $ m_V $, while the masses of $ A_S $ and the axial combination of $ A_D $ and $ \widetilde{A}_D $ are $ m_A $ as these vector resonances, given in Eq.~(\ref{Eq:vecs}), do not mix with the SM vector states. The explicit mass expressions of the additional gauge bosons are given in Ref.~\cite{BuarqueFranzosi:2016ooy}.

%%%%%%%%%%%%%%%%%%%%%%%%%%%%%%%%%%%%%%%%%%%%%%%%%%%%%%%%%%%%%%%%%%%%%%%%%%%
\subsection{The effective Lagrangian of the fermion resonances}
\label{sec: The effective theory of the fermion resonances}
%%%%%%%%%%%%%%%%%%%%%%%%%%%%%%%%%%%%%%%%%%%%%%%%%%%%%%%%%%%%%%%%%%%%%%%%%%%

In this section, we introduce the effective theory of the fermion resonances, which are important for the generation of the SM fermion masses and the Yukawa couplings of the CH. The composite fermion resonances also contribute to the effective Higgs potential and affect the generation of gravitational waves during the phase transition. 

We introduce a Dirac multiplet of fermionic resonances $ \Psi $ transforming in the antisymmetric representation of $ \SU(N)_2 $, while the composite spin-1 resonances transform in the adjoint. These fermionic resonances will mix with the elementary doublet $ q_{L,3}=(t_L,b_L)^T $ and the singlet $ t_R $ to provide the top mass. A new $ \text{U}(1)_X $ gauge symmetry is introduced to provide the correct hypercharges, embedded in the unbroken $ \SO(6)_\chi $ in Eq.~(\ref{eq: the chiral symmetry breaking}). The SM hypercharge is defined as $ Y=T_{R,1}^3+T_{R,2}^3+X $, where $ T_{R,i}^3 $ are the diagonal generators of the subgroups $ \SU(2)_{R,i} $ in $ \SP(N)_i $. The SM $ \SU(2)_L $ is the vectorial combination of the $ \SU(2)_{L,i} $ subgroups in $ \SP(N)_i $. 

In the following, we focus on the case in which both the left- and right-handed top quarks mix with top-partners in the two-index antisymmetric representation $ \mathcal{R} $ of $ \SU(N)_2 $, where the wave functions of these top-partners and their quantum number under $ \SU(N)_2\otimes \SU(6)_\chi $ in Eq.~(\ref{eq: the chiral symmetry breaking}) have the form $ QQ\chi\in (\mathcal{R},\mathbf{6}) $ as in the parentheses in Eq.~(\ref{eq: top PC-operators}). The left-handed fermionic operators of the form $ QQ\chi $ can be decomposed under the unbroken subgroup $ \text{Sp}(N)_2\otimes \SU(3)_C\otimes \text{U}(1)_X $ as \begin{equation}
\begin{aligned} 
(\mathcal{R},\mathbf{6})&=(\widetilde{\mathcal{R}},\mathbf{3},2/3)+(\widetilde{\mathcal{R}},\overline{\mathbf{3}},-2/3)+(\mathbf{1},\mathbf{3},2/3)+(\mathbf{1},\overline{\mathbf{3}},-2/3)\\ &\equiv \Psi_{\widetilde{\mathcal{R}} L}+\Psi_{\widetilde{\mathcal{R}} R}^c+\Psi_{1L}+\Psi_{1R}^c\,, \label{eq: top-partner multiplets}
\end{aligned}
\end{equation} where $ \widetilde{\mathcal{R}} $ is the two-index antisymmetric representation of the unbroken $ \text{Sp}(N)_2 $ in $ \SU(N)_2 $. Here $ \mathcal{R}=N(N-1)/2 $ and $ \widetilde{\mathcal{R}}=N(N-1)/2-1 $ representations of $ \SU(N)_2 $ and $ \SP(N)_2 $, respectively. 

For the minimal coset $ \SU(4)/\SP(4) $, the left-handed fermionic operators of the form $ QQ\chi $ can be decomposed under the unbroken subgroup $ \text{Sp}(4)_2\otimes \SU(3)_C\otimes \text{U}(1)_X $ as \begin{equation}
\begin{aligned}
(\mathbf{6},\mathbf{6})&=(\mathbf{5},\mathbf{3},2/3)+(\mathbf{5},\overline{\mathbf{3}},-2/3)+(\mathbf{1},\mathbf{3},2/3)+(\mathbf{1},\overline{\mathbf{3}},-2/3)\\ &\equiv \Psi_{5 L}+\Psi_{5 R}^c+\Psi_{1L}+\Psi_{1R}^c\,,
\end{aligned}
\end{equation} where the representations $ \widetilde{\mathcal{R}} $ and $ \mathcal{R} $ are $ \mathbf{5} $ and $ \mathbf{6} $, respectively. The explicit matrices of the composite resonances $ \Psi_{5L,5R,1L,1R} $ are given by~\cite{Dong:2020eqy} \begin{equation}
\begin{aligned} 
&\Psi_{5L}=\frac{1}{2}\left(\begin{array}{cc} T_{5L} i\sigma_2 &\sqrt{2}\widetilde{Q}_L \\ -\sqrt{2}\widetilde{Q}_L^T &  T_{5L} i\sigma_2\end{array}\right)\,, \quad \quad \widetilde{Q}_L=\left(\begin{array}{cc} T_L &X_{5/3, L} \\ B_L &  X_{2/3,L}\end{array}\right)\,, \\
&\Psi_{5R}^c=\frac{1}{2}\left(\begin{array}{cc} T_{5R}^c i\sigma_2 &\sqrt{2}\widetilde{Q}_R^c \\ -\sqrt{2}\widetilde{Q}_R^{cT} &  T_{5R}^c i\sigma_2\end{array}\right)\,, \quad \quad \widetilde{Q}_R^c=\left(\begin{array}{cc} -X_{2/3,R}^c & B_R^c \\ X_{5/3,R}^c &  -T_R^c\end{array}\right)\,, \\
&\Psi_{1L}=\frac{1}{2}\left(\begin{array}{cc} T_{1L} i\sigma_2 &0 \\ 0 & -T_{1L} i\sigma_2\end{array}\right)\,, \quad\quad \Psi_{1R}^c=\frac{1}{2}\left(\begin{array}{cc} T_{1R}^c i\sigma_2 &0 \\ 0 & -T_{1R}^c i\sigma_2\end{array}\right)\,. \label{eq: explicit embeddings top partners}
\end{aligned}
\end{equation}  Now, we consider the $ \textbf{1} $ and $ \textbf{5} $ representations of the unbroken $ \text{Sp}(4)_2 $ in $ \SU(4)_2 $. Under the decomposition $ \text{Sp}(4)_2\otimes \text{U}(1)_X \rightarrow \SU(2)_L\otimes \text{U}(1)_Y $, we get the $\SU(2)_L$ singlet $ \Psi_1 $ (denoted $ T_1 $) with $ \mathbf{1}_{2/3} $ and the fiveplet \begin{equation}
\begin{aligned}
&\left(\begin{array}{l}\textbf{5}_{2/3}\rightarrow \textbf{2}_{7/6}\oplus \textbf{2}_{1/6}\oplus \textbf{1}_{2/3}\\ \Psi_5 \rightarrow Q_X \oplus Q_5\oplus T_5 \end{array}\right)\,, \label{eq: the fiveplet}
\end{aligned}
\end{equation} where $ Q_X=(X_{5/3},X_{2/3}) $ and $ Q_5=(T,B) $. Here both $ X_{2/3}$ and $T $ receive the electric charge $+2/3$, while $X_{5/3}$ and $B$ have $+5/3$ and $-1/3$, respectively. In order to describe the linear fermion mixing in Eq.~(\ref{eq: top PC-operators}), the top quark doublet and singlet, $ q_L $ and $ t_R $, are embedded as a $ \textbf{6} $ of $ \SU(4) $ as \begin{equation}
\begin{aligned}
&\Psi_{q_L}=\frac{1}{\sqrt{2}}\left(\begin{array}{cc}0 & Q_{q_L} \\ -Q_{q_L}^T & 0 \end{array}\right)\,,  \quad  Q_{q_L}=\left(\begin{array}{cc}t_L & 0 \\ b_L & 0 \end{array}\right)\,,  \quad \Psi_{t_R}^c= \frac{t_R^c}{2}\left(\begin{array}{cc} -i\sigma_2 & 0 \\ 0 & i\sigma_2 \end{array}\right)\,.
\end{aligned}
\end{equation} 

For non-minimal cosets $ \SU(N)/\SP(N) $, the explicit matrices of $ \Psi_{\widetilde{\mathcal{R}}L,\widetilde{\mathcal{R}}R,1L,1R} $ in Eq.~(\ref{eq: top-partner multiplets}) are given by\begin{equation}
\begin{aligned}
&\Psi_{\widetilde{\mathcal{R}}L}=\frac{1}{2}\left(\begin{array}{ccccc} A_{1L} i\sigma_2 &\sqrt{2}\widetilde{Q}_L & \sqrt{2}\widetilde{Q}_{L}^{1,3} & \cdots &  \sqrt{2}\widetilde{Q}_{L}^{1,N/2}  \\ -\sqrt{2}\widetilde{Q}_L^T &  A_{2L} i\sigma_2 & \sqrt{2}\widetilde{Q}_{L}^{2,3} & \cdots &  \sqrt{2}\widetilde{Q}_{L}^{2,N/2}  \\-\sqrt{2}(\widetilde{Q}_L^{1,3})^T &  -\sqrt{2}(\widetilde{Q}_L^{2,3})^T & A_{3L} i\sigma_2 & \cdots &  \sqrt{2}\widetilde{Q}_{L}^{3,N/2} \\ \vdots & \vdots & \vdots & \ddots & \vdots \\  -\sqrt{2}(\widetilde{Q}_L^{1,N/2})^T &  -\sqrt{2}(\widetilde{Q}_L^{2,N/2})^T &  -\sqrt{2}(\widetilde{Q}_L^{3,N/2})^T & \cdots & A_{N/2 L} i\sigma_2 \end{array}\right)\,, \\& 
 \Psi_{\widetilde{\mathcal{R}}R}^c=\frac{1}{2}\left(\begin{array}{ccccc} A_{1R}^c i\sigma_2 &\sqrt{2}\widetilde{Q}_R^c & \sqrt{2}\widetilde{Q}_{R}^{c1,3} & \cdots &  \sqrt{2}\widetilde{Q}_{R}^{c1,N/2}  
 \\ -\sqrt{2}\widetilde{Q}_R^{cT} &  A_{2R}^c i\sigma_2 & \sqrt{2}\widetilde{Q}_{R}^{c2,3} & \cdots &  \sqrt{2}\widetilde{Q}_{R}^{c2,N/2}
 \\-\sqrt{2}(\widetilde{Q}_R^{c1,3})^T &  -\sqrt{2}(\widetilde{Q}_R^{c2,3})^T & A_{3R}^c i\sigma_2 & \cdots &  \sqrt{2}\widetilde{Q}_{R}^{c3,N/2} 
 \\ \vdots & \vdots & \vdots & \ddots & \vdots 
 \\  -\sqrt{2}(\widetilde{Q}_R^{c1,N/2})^T &  -\sqrt{2}(\widetilde{Q}_R^{c2,N/2})^T &  -\sqrt{2}(\widetilde{Q}_R^{c3,N/2})^T & \cdots & A_{N/2 R}^c i\sigma_2 \end{array}\right)\,,
\\
&\Psi_{1L}=\frac{1}{2}T_{1L} \Sigma_{Q0} \,, \quad\quad \Psi_{1R}^c=\frac{1}{2}T_{1R}^c \Sigma_{Q0}\,, \label{eq: explicit embeddings top partners general}
\end{aligned}
\end{equation} where $ \Sigma_{Q0} $ is the EW unbroken vacuum matrix given by Eq.~(\ref{eq: EW unbroken vacuum}) and \begin{equation}
\begin{aligned}
 &\widetilde{Q}_L=\left(\begin{array}{cc} T_L &X_{5/3, L} \\ B_L &  X_{2/3,L}\end{array}\right)\,, \quad \quad \widetilde{Q}_R^c=\left(\begin{array}{cc} -X_{2/3,R}^c & B_R^c \\ X_{5/3,R}^c &  -T_R^c\end{array}\right)\,, \\
  &\widetilde{Q}_L^{i,j}=\left(\begin{array}{cc} T_L^{i,j} &X_{5/3, L}^{i,j} \\ B_L^{i,j} &  X_{2/3,L}^{i,j}\end{array}\right)\,, \quad \quad \widetilde{Q}_R^{ci,j}=\left(\begin{array}{cc} -X_{2/3,R}^{ci,j} & B_R^{ci,j} \\ X_{5/3,R}^{ci,j} &  -T_R^{ci,j}\end{array}\right)\,, \\
& A_1 = T_5 +\sum_{I=3}^{N/2} \frac{1}{\sqrt{I}} T_{I(2I-1)-1}\,,  \quad \quad A_2 = T_5 - \sum_{I=3}^{N/2}  \frac{1}{\sqrt{I}} T_{I(2I-1)-1}\,, \\
& A_{i\geq 3} = \frac{i-1}{\sqrt{i}}T_{i(2i-1)-1}+\sum_{I=i+1}^{N/2}\frac{1}{\sqrt{I}}T_{I(2I-1)-1}\,. 
\end{aligned}
\end{equation} In order to describe the linear fermion mixing in Eq.~(\ref{eq: top PC-operators}), the top quark doublet and singlet, $ q_L $ and $ t_R $, are embedded as a $ \textbf{6} $ of $ \SU(4) \subseteq \SU(N) $ as \begin{equation}
\begin{aligned}
&\Psi_{q_L}=\frac{1}{\sqrt{2}}\left(\begin{array}{ccccc}0 & Q_{q_L} & 0 & \cdots & 0 \\ -Q_{q_L}^T & 0 & 0 & \cdots & 0 \\ 0 & 0 & 0 & \cdots & 0 \\ \vdots & \vdots & \vdots & \ddots & \vdots \\ 0 & 0 & 0 & \cdots & 0 \end{array}\right)\,, \quad \quad  Q_{q_L}=\left(\begin{array}{ccccc}t_L & 0 & 0 & \cdots & 0 \\ b_L & 0 & 0 & \cdots & 0 \\ 0 & 0 & 0 & \cdots & 0 \\ \vdots & \vdots & \vdots & \ddots & \vdots \\ 0 & 0 & 0 & \cdots & 0 \end{array}\right)
\,, \quad \quad \\ & \Psi_{t_R}^c=\frac{t_R^c}{2} \text{Diag}(-i\sigma_2,i\sigma_2,0,0,\dots)\,.
\end{aligned}
\end{equation} 

Above we have considered the $ \textbf{1} $ and $ \widetilde{\mathcal{R}}=N(N-1)/2-1 $ representations of the unbroken $ \text{Sp}(N)_2 $ in $ \SU(N)_2 $. For example, if we consider the CH models with coset $ \SU(6)/\SP(6) $ or $ \SU(8)/\SP(8) $, we have the $ \textbf{1} $ and $ \widetilde{\mathcal{R}}=\textbf{14} $ or $ \textbf{27} $ representations of the unbroken $ \text{Sp}(6)_2 $ or $ \text{Sp}(8)_2 $, respectively. Here the top-partner singlet $ \textbf{1} $ is identified by $ T_1 $ in Eq.~(\ref{eq: explicit embeddings top partners general}). Under the decomposition $ \text{Sp}(6)_2\otimes \text{U}(1)_X \rightarrow \SP(4)_2\otimes \SU(2)_3 \otimes \text{U}(1)_X $ or $ \text{Sp}(8)_2\otimes \text{U}(1)_X \rightarrow \SP(6)_2\otimes \SU(2)_4 \otimes \text{U}(1)_X $, we get \begin{equation}
\begin{aligned}
&\textbf{14}_{2/3}\rightarrow (\textbf{1},\textbf{1})_{3/2}+(\textbf{6},\textbf{1})_{3/2}\oplus (\textbf{4},\textbf{2})_{2/3}\,, \\
&\textbf{27}_{2/3}\rightarrow (\textbf{1},\textbf{1})_{3/2}+(\textbf{14},\textbf{1})_{3/2}\oplus (\textbf{6},\textbf{2})_{2/3}\,,
 \label{eq: multiplet SU(6) and SU(8)}
\end{aligned}
\end{equation} respectively. The first term represents the top-partner singlets $ T_{14} $ for $ \SU(6)/\SP(6) $, $ T_{27} $ for $ \SU(8)/\SP(8) $ and $ T_{N(N-1)/2-1} $ for $ \SU(N)/\SP(N) $, while the second term contains the top-partners in the more minimal CH models. Finally, the third term contains $ \widetilde{Q}_L^{i,3} $ for $ \SU(6)/\SP(6) $, $ \widetilde{Q}_L^{i,4} $ for $ \SU(8)/\SP(8) $ and $ \widetilde{Q}_L^{i,N/2} $ for $ \SU(N)/\SP(N) $ with $ i=1,\dots,N/2-1 $, see the various top-partners in Eq.~(\ref{eq: explicit embeddings top partners general}).   

To leading order, the effective Lagrangian for the elementary and composite fermions can then be decomposed into three parts
 \begin{equation}
\begin{aligned}
\mathcal{L}_{\rm ferm}=\mathcal{L}_{\rm elem}+\mathcal{L}_{\rm comp}+\mathcal{L}_{\rm mix}\,. \label{eq: fermion resonances lag}
\end{aligned} 
\end{equation} The Lagrangian for the elementary fermions is \begin{equation}
\begin{aligned}
\mathcal{L}_{\rm elem}=& i \overline{q}_{L,3} \slashed{D} q_{L,3} + i \overline{t}_R \slashed{D}t_R \,,
\end{aligned} 
\end{equation} where the covariant derivatives is given by \begin{equation}
\begin{aligned}
D_\mu q_{L,3} = &\left(\partial_\mu - ig W_\mu^i \frac{\sigma_i}{2}-i\frac{1}{6}g^\prime B_\mu -ig_S G_\mu\right)q_{L,3}\,, \\
D_\mu t_L = &\left(\partial_\mu -i\frac{2}{3}g^\prime B_\mu -ig_S G_\mu\right)t_L
\,.
\end{aligned} 
\end{equation} The Lagrangian for the composite fermions containing their gauge-kinetic, mass and interaction terms between the $ \widetilde{\mathcal{R}} $-plet and singlet can be written as~\cite{Panico:2015jxa,Dong:2020eqy} \begin{equation}
\begin{aligned}
\mathcal{L}_{\rm comp}=& i \text{Tr}[\Psi_{\widetilde{\mathcal{R}}L}^\dagger \overline{\sigma}^\mu D_\mu \Psi_{\widetilde{\mathcal{R}}L}]+i \text{Tr}[(\Psi_{\widetilde{\mathcal{R}}R}^c)^\dagger \sigma^\mu D_\mu \Psi_{\widetilde{\mathcal{R}}R}^c]\\&+i \text{Tr}[\Psi_{1L}^\dagger \overline{\sigma}^\mu D_\mu \Psi_{1L}]+i \text{Tr}[(\Psi_{1R}^c)^\dagger \sigma^\mu D_\mu \Psi_{1R}^c]\\&-\left(m_{\widetilde{\mathcal{R}}}\text{Tr}[\Psi_{\widetilde{\mathcal{R}}L}\Sigma_{Q0}\Psi_{\widetilde{\mathcal{R}}R}^c\Sigma_{Q0}]+m_1\text{Tr}[\Psi_{1L}\Sigma_{Q0}\Psi_{1R}^c\Sigma_{Q0}]+\text{h.c.}\right)\\&-\left(ic_L\text{Tr}[\Psi_{\widetilde{\mathcal{R}}L}^\dagger \overline{\sigma}^\mu d_{2,\mu} \Psi_{1L}]+ic_R\text{Tr}[\Psi_{\widetilde{\mathcal{R}}R}^\dagger \sigma^\mu d_{2,\mu} \Psi_{1R}]+\text{h.c.}\right)
\,, \label{eq: comp lagrangian}
\end{aligned} 
\end{equation} where the covariant derivatives are given by \begin{equation}
\begin{aligned}
D_\mu \Psi_{\widetilde{\mathcal{R}}} =&\left(\partial_\mu - i\frac{2}{3}g^\prime B_\mu T_R^3 -ie_{2,\mu} - ig_S G_\mu \right)\Psi_{\widetilde{\mathcal{R}}}\,, \\
D_\mu \Psi_1 =&\left(\partial_\mu -i \frac{2}{3}g^\prime B_\mu  T_R^3  - ig_S G_\mu  \right)\Psi_1\,,
\end{aligned} 
\end{equation} and the $ d $ and $ e $ symbols are defined in Eq.~(\ref{eq: d and e symbols vec res}) which contain the heavy spin-1 resonances. Finally, the mixing terms between the SM fermions, composite PNGBs and fermion resonances invariant under $ \SU(N) $ are given by~\cite{Dong:2020eqy} \begin{equation}
\begin{aligned}
\mathcal{L}_{\rm mix}=&-y_{L\widetilde{\mathcal{R}}}f\text{Tr}[\Psi_{q_L}U_Q\Psi_{\widetilde{\mathcal{R}}R}^c U_Q^T]-y_{L1}f\text{Tr}[\Psi_{q_L}U_Q\Psi_{1R}^c U_Q^T]\\&-y_{R\widetilde{\mathcal{R}}}f\text{Tr}[\Psi_{t_R}^c U_Q\Psi_{\widetilde{\mathcal{R}}L}U_Q^T]-y_{R1}f\text{Tr}[\Psi_{t_R}^c U_Q\Psi_{1L} U_Q^T]+\text{h.c.}\,.
\end{aligned}
\end{equation}

Finally, from the effective Lagrangian $ \mathcal{L}_{\rm ferm} $ in Eq.~(\ref{eq: fermion resonances lag}), the mass and mixing terms of the top quark and the top-partners (given in Eq.~(\ref{eq: explicit embeddings top partners})) with $ +2/3 $ charge are given by $ \overline{\Psi}_{2/3L} M_{\Psi_{2/3}}\Psi_{2/3R} $ with \begin{equation}
\begin{aligned}
 \Psi_{2/3} = \left( t\,,\, T_{1} \,,\, T_{1'} \,,\, T \,,\, X_{2/3}  \,,\, T_{5} \,,\, T_{14} \,,\, T_{27} \,,\, \dots \,,\, T_{N(N-1)/2-1} \right)\,, \label{eq: top partner 2/3 charge}
\end{aligned}
\end{equation} where the mass matrix $ M_{\Psi_{2/3}} $ of the top quark and top-partners with $ +2/3 $ charge has the form  \begin{equation}
\begin{aligned}
 \left(\begin{array}{ccccccccccc} 0 & \frac{y_{L1}fs_\theta}{\sqrt{2}}  & \frac{y_{L1'}fs_\theta}{\sqrt{2}} & y_{L\widetilde{\mathcal{R}}}f c_{\theta/2}^2 & y_{L\widetilde{\mathcal{R}}}fs_{\theta/2}^2 
 & 0 &  \frac{y_{L\widetilde{\mathcal{R}}}fs_\theta}{\sqrt{6}} & \frac{y_{L\widetilde{\mathcal{R}}}fs_\theta}{\sqrt{8}} & \cdots & \frac{y_{L\widetilde{\mathcal{R}}}fs_\theta}{\sqrt{2 N}} \\ y_{R1}fc_\theta & -m_1   & 0 & 0 & 0 & 0  & 0 & 0 & \cdots & 0 \\ y_{R1'}fc_\theta & 0 & -m_{1'}  & 0 & 0 & 0 & 0 & 0 & \cdots & 0 \\ -\frac{y_{R\widetilde{\mathcal{R}}}fs_\theta}{\sqrt{2}} & 0 & 0 & -m_{\widetilde{\mathcal{R}}}  & 0 & 0 & 0 & 0 & \cdots & 0 \\ \frac{y_{R\widetilde{\mathcal{R}}}fs_\theta}{\sqrt{2}} & 0 & 0 & 0 & -m_{\widetilde{\mathcal{R}}} & 0 & 0 & 0 & \cdots & 0 
 \\
0  & 0 & 0 & 0 & 0  & -m_5 & 0 & 0 & \cdots & 0  
\\ 
\frac{y_{R\widetilde{\mathcal{R}}}fc_\theta}{3}   & 0 & 0 & 0 & 0  & 0 & -\frac{m_{\widetilde{\mathcal{R}}}}{3} & 0 & \cdots & 0  
\\ \frac{y_{R\widetilde{\mathcal{R}}}fc_\theta}{4}   & 0 & 0 & 0 & 0 & 0 & 0 & -\frac{m_{\widetilde{\mathcal{R}}}}{4} & \cdots & 0 
\\ \vdots  & \vdots & \vdots & \vdots &\vdots & \vdots  & \vdots & \vdots  & \ddots & \vdots  
\\ \frac{y_{R\widetilde{\mathcal{R}}}fc_\theta}{N}  &  0 & 0 & 0 & 0 & 0 & 0 & 0 & \cdots & -\frac{m_{\widetilde{\mathcal{R}}}}{N} \end{array}\right) \,.  \label{eq: top mass matrix}
\end{aligned}
\end{equation} By diagonalizing this mass matrix and expanding in $ s_\theta $, we then find the top mass: \begin{equation}
\begin{aligned}
m_{\rm top}= & \frac{v_{\rm SM}}{\sqrt{2}}\frac{m_1 m_{\widetilde{\mathcal{R}}}}{\sqrt{m_1^2 + y_{R1}^2f^2}\sqrt{m_{\widetilde{\mathcal{R}}}^2 + y_{L\widetilde{\mathcal{R}}}^2f^2}}\times\\ & \Bigg\vert  \frac{y_{L1}y_{R1}f}{m_1}+\frac{y_{L1'}y_{R1'}f}{m_{1'}}-\left(1+\sum_{i=3}^{N/2}\frac{1}{\sqrt{i}}\right)\frac{y_{L\widetilde{\mathcal{R}}}y_{R\widetilde{\mathcal{R}}}f}{m_{\widetilde{\mathcal{R}}}}\Bigg\vert + \mathcal{O}(s_\theta^2) \,,
\label{eq: top mass}
\end{aligned}
\end{equation} where $ T_5 $ is the only $ 2/3 $ charged top-partner that does not mix with the top quark and, therefore, does not contribute to the top mass.  

\section{The scalar potential at zero and finite temperature}
\label{sec: The scalar potential and the condition of SFOEWPT}

The zero temperature effective scalar potential of $ h $ and $ \eta $ fields is generally expected to be of the form 
\beq \label{eq: pot of h eta}
V(h,\eta)=\frac{\mu_h^2}{2}h^2+\frac{\lambda_h}{4}h^4+\frac{\mu_\eta^2}{2}\eta^2+\frac{\lambda_\eta}{4}\eta^4+\frac{\lambda_{h\eta}}{2}h^2\eta^2\,, 
\eeq 
while the Higgs kinetic term in Eq.~(\ref{eq:effLag kin}) is canonically normalized by making the replacement  \beq h \rightarrow v_{\rm SM} +h \cos\theta \,,
\eeq resulting in the physical masses  \beq \label{eq: Higgs and eta mass expressions}
m_h^2= -2\mu_h^2 \cos^2 \theta \,, \quad \quad m_\eta^2=\mu_\eta^2+\lambda_{h\eta}v_{\rm SM}^2\,,
\eeq where $\theta$ is the vacuum misalignment angle as defined in Eq.~(\ref{SM VEV}). In the following sections, we will match the coefficients in Eq.~\eqref{eq: pot of h eta} 
with the underlying composite dynamics

The zero-temperature potential has four non-trivial extrema at \beq \label{eq: VEV expressions}
(v_h,0) = \left( \pm\sqrt{-\frac{\mu_h^2}{\lambda_h}} ,0\right) \quad {\textrm{and}}\quad (0,v_\eta) =\left(0, \pm\sqrt{-\frac{\mu_\eta^2}{\lambda_\eta}}\right)
\eeq where $ v_h \equiv v_{\rm SM}= f \sin \theta $
as required by Eqs.~(\ref{eq: WZ masses and SM VEV}) and~(\ref{SM VEV}) to yield correct electroweak boson masses. To be extrema of the potential, the points $ (v_h,0) $ and $ (0,v_\eta) $ need to satisfy the inequalities \begin{equation} \begin{aligned}\label{eq: inequalities 1}
&\mu^2_h < 0\,, \quad \quad \lambda_h >0\,, \quad \quad \mu^2_\eta < 0\,, \quad \quad \lambda_\eta >0\,,
\end{aligned}\end{equation} and, to ensure that they are local minima, the inequalities coming 
from the Hessian matrix \begin{equation} \begin{aligned}\label{eq: inequalities 2}
&\lambda_h\mu_\eta^2  > \lambda_{h\eta}\mu_h^2 \,, \quad \quad \lambda_\eta \mu_h^2  > \lambda_{h\eta}\mu_\eta^2.
\end{aligned}\end{equation}  Furthermore, the minimum $ (v_h,0) $ along the $ h $ direction breaking the EW symmetry should be the true vacuum, which can be ensured by the inequality  \begin{equation} \begin{aligned}\label{eq: inequalities 3}
& V(v_h,0) < V(0,v_\eta) \quad \Rightarrow \quad \mu_\eta^2 \sqrt{\lambda_h} >\mu_h^2\sqrt{\lambda_\eta}\,.
\end{aligned}\end{equation}

At finite temperature $T$, the potential receives thermal corrections which affect the vacuum structure. To leading order in $ T^2 $, the finite temperature potential is given by~\cite{Espinosa:2011ax} \beq \label{eq: finite temperature potential}
V_T(h,\eta)=\frac{\mu_h^2+c_h T^2}{2}h^2+\frac{\lambda_h}{4}h^4+\frac{\mu_\eta^2+c_\eta T^2}{2}\eta^2+\frac{\lambda_\eta}{4}\eta^4+\frac{\lambda_{h\eta}}{2}h^2\eta^2\,,
\eeq where \beq 
c_h=\frac{3g^2+g^{\prime 2}}{16}+\frac{y_t^2}{4}+\frac{\lambda_h}{2}+\frac{\lambda_{h\eta}}{12}\,, \quad\quad c_\eta= \frac{\lambda_\eta}{4}+\frac{\lambda_{h\eta}}{3}\,.
\eeq 

The necessary condition for SFOEWPT is the appearance of two degenerate vacua at some
critical temperature $ T_c $. In our model, this is realized due to a two-step phase transition\footnote{We will here use the method similar to those in Refs~\cite{Bian:2018bxr,Bian:2018mkl,Bian:2019kmg,DeCurtis:2019rxl,Xie:2020bkl}}, where the background values $ (\langle h \rangle(T), \langle \eta \rangle(T)) $ of the fields $ h $ and $ \eta $ developed as $ (0,0)\rightarrow (0,v_\eta) \rightarrow (v_{h},0) $ as the universe cooled down from $ T\gg m_h $ to $ T\approx 0 $.
At some critical temperature $ T_c $, according to Eq.~(\ref{eq: finite temperature potential}), there exist two degenerate vacua, which are given by \begin{equation} \begin{aligned}\label{eq: degenerate vacua}
v_{h,c} =\sqrt{-\frac{\mu_h^2+c_hT_c^2}{\lambda_h}}\,, \quad \quad v_{\eta,c} =\sqrt{-\frac{\mu_\eta^2+c_\eta T_c^2}{\lambda_\eta}}
\end{aligned}\end{equation} satisfying the inequalities \begin{equation} \begin{aligned}\label{eq: inequalities 4}
&\mu_h^2 +c_h T_c^2 < 0\,, \quad \quad  \lambda_h(\mu_\eta^2+c_\eta T_c^2)> \lambda_{h\eta}(\mu_h^2 + c_h T_c^2)\,, \\ &  \mu_\eta^2 +c_\eta T_c^2 < 0\,, \quad\quad \lambda_\eta(\mu_h^2+c_h T_c^2)> \lambda_{h\eta}(\mu_\eta^2 + c_\eta T_c^2)\,.
\end{aligned}\end{equation} This critical temperature can be found by solving the equation $ V_{T_c}(v_{h,c},0)=V_{T_c}(0,v_{\eta,c}) $, which gives rise to\begin{equation} \begin{aligned}\label{eq: critical temperature}
T_c^2 = \frac{\mu_\eta^2\sqrt{\lambda_h}-\mu_h^2\sqrt{\lambda_\eta}}{c_h\sqrt{\lambda_\eta}-c_\eta\sqrt{\lambda_h}}\,.  
\end{aligned}\end{equation} Assuming that the value of $ T_c $ is real and the inequality in Eq.~(\ref{eq: inequalities 3}) is satisfied, we obtain the inequality $ c_h\sqrt{\lambda_\eta}>c_\eta \sqrt{\lambda_h} $. Inserting the expression of $ T_c $ into the inequalities in Eq.~(\ref{eq: inequalities 4}), we find that \begin{equation} \begin{aligned}\label{eq: inequalities 5}
\frac{c_\eta}{c_h}<\frac{\mu_\eta^2}{\mu_h^2}\,.
\end{aligned}\end{equation} Combining the inequality in Eq.~(\ref{eq: inequalities 5}) with the inequalities in Eqs.~(\ref{eq: inequalities 1})-(\ref{eq: inequalities 3}), the condition of two degenerate vacua for $ V_T(h,\eta) $ in Eq.~(\ref{eq: finite temperature potential}) is fullfilled when \begin{equation} \begin{aligned}\label{eq: degenerate vacuums condition}
\frac{c_\eta}{c_h}<\frac{\mu_\eta^2}{\mu_h^2}<\frac{\sqrt{\lambda_\eta}}{\sqrt{\lambda_h}}<\frac{\lambda_{h\eta}}{\lambda_h}\,.
\end{aligned}\end{equation} 

Note that degenerate vacua are necessary, but not sufficient for a first-order electroweak phase transition (FOEWPT). To achieve a FOEWPT, the critical condition \begin{equation} \begin{aligned}\label{eq: critical condition}
\frac{S_3(T_n)}{T_n}\sim 4\ln \left(\frac{\xi m_{\rm Planck}}{T_n}\right)\sim 140
\end{aligned}\end{equation} should be satisfied at some nucleation temperature $ T_n $, which we obtain by calculating the bubble nucleation rate per volume in the early universe \begin{equation} \begin{aligned}\label{}
\frac{\Gamma}{V}	\approx T^4 \left(\frac{S_3}{2\pi T}\right)^{3/2}e^{-S_3(T)/T}\,.
\end{aligned}\end{equation} Here $ S_3 $ is the classical action of the $ O(3) $ symmetric bounce solution~\cite{Linde:1981zj} and $ \xi\approx 0.03 $. 

In the following, we parametrise the strength of the
transition by the quantity $ v_{h,c}/T_c $, where $ v_c $ and $ T_c $ are given by Eqs.~(\ref{eq: degenerate vacua}) and~(\ref{eq: critical temperature}), respectively. According to Ref.~\cite{Curtin:2014jma}, a value of \begin{equation} \begin{aligned}\label{eq: strong criterion}
\frac{v_{h,c}}{T_c}>1.6
\end{aligned}\end{equation} suggests that electroweak baryogenesis may be efficient. This efficiency arises from the prevention of sphalerons from washing out the baryon asymmetry within the broken phase. Moreover, it indicates a strong first-order transition. Therefore, in the following numerical calculations, this criterion will be fulfilled resulting in a strong first-order electroweak phase transitions (SFOEWPT). 

In the above we have derived the conditions that ensure SFOEWPT using the mean field approximation \eqref{eq: finite temperature potential}, for clarity of presentation. However, in the following numerical analysis we include the finite temperature corrections beyond the mean field approximation, using the $\textsf{CosmoTransitions}$ package~\cite{Wainwright:2011kj}, where the thermal corrections are implemented as
\begin{equation}
    \Delta V(T) = \sum_i \frac{T^4}{2\pi^2}\left( - \sum_{k=1}^8 \frac{x_i^2}{k^2} K_2(kx_i) - \sum_{k=1}^8 \frac{(-1)^kx_i^2}{k^2} K_2(kx_i) \right),
\end{equation}
where $K_2$ is the modified Bessel function of the second kind, $x_i=m_i/T$ and the sum runs over the scalar particles, $m_i = m_h,m_\eta$. 

In the following subsection, we investigate the various contributions to the parameters, $ \mu_{h,\eta} $ and $ \lambda_{h,\eta,h\eta} $, in the scalar potential given in Eq.~(\ref{eq: pot of h eta}) from the spin-1 resonances, fermion resonances and vector-like masses of the hyper-fermions. 

\subsection{Contributions from the spin-1 resonances}

After integrating out the heavy spin-1 resonances in Eq.~(\ref{eq: VA lag}), we obtain an effective Lagrangian 
for gauge fields in the unitary gauge given by \begin{equation}
\begin{aligned} \label{eq: eff lag gauge}
\mathcal{L}_{\rm eff}^{\rm gauge}=\frac{1}{2}P_T^{\mu\nu}\Bigg(&\frac{\Pi_0(p^2)}{2}\text{Tr}[\widetilde{A}_\mu \widetilde{A}_\nu]-p^2(W_\mu^a W_\nu^a + B_\mu B_\nu)\\ &+\frac{\Pi_1(p^2)}{4}\text{Tr}\big[(\widetilde{A}_\mu \Sigma+\Sigma \widetilde{A}_\mu^T)(\widetilde{A}_\nu \Sigma+\Sigma \widetilde{A}_\nu^T)^\dagger\big]\Bigg)
\end{aligned}  
\end{equation} with $ \widetilde{A}_\mu \equiv gW_\mu^a T_L^a+g^\prime B_\mu T_R^3 $. The transverse and longitudinal projection operators are defined as \begin{equation}
\begin{aligned}
P_T^{\mu\nu}=g^{\mu\nu}-\frac{p^\mu p^\nu}{p^2}\,, \quad \quad \quad P_L^{\mu\nu}=\frac{p^\mu p^\nu }{p^2}\,,
\label{eq: projection operators}
\end{aligned}
\end{equation} and the form factors are given by the Weinberg sum rules~\cite{Bian:2019kmg} \begin{equation}
\begin{aligned} 
\Pi_0(p^2)=\frac{ p^2 f_K^2}{p^2-m_V^2}\,, \quad \quad \quad \Pi_1(p^2)=\frac{ f^2 m_V^2 m_A^2}{(p^2-m_V^2)(p^2-m_A^2)}\,.
\end{aligned}  
\end{equation} From the definition of the Fermi decay constant, it is obtained that $ f=\sqrt{f_1^2-r^2 f_2^2} $~\cite{BuarqueFranzosi:2016ooy}, where $ r $ is defined in Eq.~(\ref{eq: VA lag}). By assuming the lightest axial and vector resonance multiplets saturate the absorptive part of the vacuum polarizations, another set of useful sum rules appears, given by~\cite{Bian:2019kmg} \begin{equation}\begin{aligned} f_K^2 = \frac{f^2}{2}+f_1^2 \,, \quad \quad \quad f_K^2m_V^2=f_1^2 m_A^2\,. \label{eq: sum rule gauge 2} \end{aligned}  \end{equation}

From the effective Lagrangian in Eq. (\ref{eq: eff lag gauge}), at the one-loop level of the gauge interactions, we can write the Coleman-Weinberg potential in Euclidean momenta space  \begin{equation}
\begin{aligned} 
V_{\rm gauge}^{\rm IR}=\frac{3}{2}\int \frac{d^4 p_E}{(2\pi)^4}\Bigg(&2\ln \left[\Pi^W_0 +\frac{1}{4}\Pi_1 \left(\frac{h}{\pi_Q}\right)^2\sin^2\left(\frac{\pi_Q}{f}\right)\right]\\ &+\ln \left[\Pi_0^B\Pi_0^W+\frac{1}{2}\Pi_1\left(\frac{h}{\pi_Q}\right)^2\sin^2\left(\frac{\pi_Q}{f}\right)(\Pi_0^B+\Pi_0^W)\right] \Bigg) \,, \label{eq: pot from EW gauge int}
\end{aligned}  
\end{equation} where $ \pi_Q\equiv \sqrt{h^2+\eta^2} $. Furthermore, we have \begin{equation}
\begin{aligned} 
\Pi^B_0(p^2)=-\frac{p^2}{g^{\prime 2}}+N_R \Pi_0(p^2)\frac{g^{\prime 2}}{g^2}\,, \quad \quad \Pi^W_0(p^2)=-\frac{p^2}{g^{2}}+N_L \Pi_0(p^2)\,,
\end{aligned}  
\end{equation} where $N_{L,R}$ are, respectively, the number of $\SU(2)_{L,R}$ hyper-fermion doublets in the model. By expanding $ s_Q $, this potential can be written as {\setlength{\thickmuskip}{.1\thickmuskip}
  \setlength{\medmuskip}{0.9\medmuskip}\begin{equation}
\begin{aligned} 
V_{\rm gauge}^{\rm IR}(h)\approx\frac{6}{2}\int \frac{d^4 p_E}{(2\pi)^4} \ln\left[1+\frac{\Pi_1}{4\Pi^W_0}\frac{h^2}{f^2}\right]+\frac{3}{2}\int \frac{d^4 p_E}{(2\pi)^4} \ln\left[1+\left(\frac{\Pi_1}{4\Pi^B_0}+\frac{\Pi_1}{4\Pi^W_0}\right)\frac{h^2}{f^2}\right] \,,
\end{aligned}  
\end{equation}} By expanding the above potential up to $ h^4 $, we can match it to Eq.~(\ref{eq: pot of h eta}), which yields\begin{equation}
\begin{aligned} 
(\mu_h^2)^{\rm IR}_g&=\frac{3}{4f^2}\int \frac{d^4 p_E}{(2\pi)^4} \left(\frac{\Pi_1}{\Pi^B_0}+3\frac{\Pi_1}{\Pi^W_0}\right) \,, \\
(\lambda_h)^{\rm IR}_g&=-\frac{3}{16f^4}\int \frac{d^4 p_E}{(2\pi)^4} \left[2\left(\frac{\Pi_1}{\Pi^W_0}\right)+\left(\frac{\Pi_1}{\Pi^B_0}+\frac{\Pi_1}{\Pi^W_0}\right)^2\right]\,, \\
(\mu_\eta^2)^{\rm IR}_g &=(\lambda_\eta)^{\rm IR}_g=(\lambda_{h \eta})^{\rm IR}_g=0 \,.
\end{aligned}  
\end{equation}

The potential contributions from the higher order (UV) operators can be written down using the spurion trick~\cite{Alanne:2018wtp}. To next-to-leading order of contributions from the spin-1 resonances, the effective potential can be written as\begin{equation}
\begin{aligned} 
V_{\rm gauge}^{\rm UV}=&c_g f^4  g^2  \text{Tr}[T_L^a \Sigma (T_L^a)^T \Sigma^\dagger]+c_{g^{\prime}} f^4 g^{\prime 2}\text{Tr}[Y \Sigma Y^T \Sigma^\dagger] \\ &
+\frac{d_g}{(4\pi)^2} f^4 g^4 \Big(\text{Tr}[T_L^a \Sigma (T_L^a)^T \Sigma^\dagger]^2+\text{Tr}[T_L^aT_L^a T_L^b \Sigma (T_L^b)^T \Sigma^\dagger]\\ & \quad\quad\quad +\text{Tr}[T_L^aT_L^a  \Sigma (T_L^b)^T (T_L^b)^T   \Sigma^\dagger]+\text{Tr}[T_L^a \Sigma (T_L^a)^T   \Sigma^\dagger T_L^b \Sigma (T_L^b)^T \Sigma^\dagger]\Big) \\ &
+\frac{d_{g^\prime}}{(4\pi)^2} f^4 g^{\prime 4} \Big(\text{Tr}[Y \Sigma Y^T \Sigma^\dagger]^2+\text{Tr}[Y Y Y  \Sigma Y^T \Sigma^\dagger]\\ & \quad\quad\quad +\text{Tr}[YY  \Sigma Y^T Y^T   \Sigma^\dagger]+\text{Tr}[Y \Sigma Y^T   \Sigma^\dagger Y \Sigma Y^T \Sigma^\dagger]\Big) + \dots\,, \label{eq: pot from EW gauge int UV}
\end{aligned}  
\end{equation} where the matrices for $T_L^a $ and $ Y \equiv T_R^3$ are given in Eq.~(\ref{eq: SU(2)L and U(1)Y matrices}), while $c_{g,g'}$ and $d_{g,g'}$ are non-perturbative coefficients which can be determined by the lattice simulations. From this effective potential, the contributions to the scalar potential in Eq.~(\ref{eq: pot of h eta}) are then \begin{equation}
\begin{aligned} 
(\mu_h^2)^{\rm UV}_g&=3 c_g f^2 g^2 + c_{g^\prime} f^2 g^{\prime 2} -\frac{27+18(N_L-1)}{32\pi^2 }d_g f^2g^4 -\frac{3+2(N_R-1)}{32\pi^2 }d_{g^\prime} f^2g^{\prime 4}\,, \\
(\lambda_h)^{\rm UV}_g&= \frac{27}{32\pi^2 }d_g g^4+\frac{3}{32\pi^2 }d_{g^{\prime}} g^{\prime 4}\,, \quad \quad 
(\mu_\eta^2)^{\rm UV}_g=(\lambda_\eta)^{\rm UV}_g=(\lambda_{h \eta})^{\rm UV}_g=0\,.
\end{aligned}  
\end{equation} \phantom{ }

\subsection{Contributions from the fermion resonances}
\label{eq: Contributions from the fermion resonances}

After integrating out the heavy top-partners in Eq.~(\ref{eq: fermion resonances lag}), like we did for the heavy vector resonances above, we obtain an effective Lagrangian for the top, 
\begin{equation}
\begin{aligned}
\mathcal{L}_{\rm eff}^{\rm top}=&\Pi_0^q (p)\text{Tr}\left[\overline{\Psi}_{q_L}\slashed{p}\Psi_{q_L}\right]+\Pi_0^t (p)\text{Tr}\left[\overline{\Psi}_{t_R}^c \slashed{p}\Psi_{t_R}^c\right]\\ & \Pi_1^q (p)\text{Tr}\left[\Psi_{q_L}\Sigma_Q^*\slashed{p}\Psi_{q_L}^c \Sigma_Q^*\right]+\Pi_1^t (p)\text{Tr}\left[\Psi_{t_R}\Sigma_Q^*\slashed{p}\Psi_{t_R}^c \Sigma_Q^*\right] \\ &+M_1^t (p)\text{Tr}\left[\Psi_{q_L}\Sigma_Q^*\Psi_{t_R}^c \Sigma_Q^*\right]+\text{h.c.}\,,
\label{eq: top eff lag}
\end{aligned}
\end{equation} where the form factors are given by the Weinberg sum rules~\cite{Bian:2019kmg} \begin{equation}
\begin{aligned}
&\Pi_0^q(p)=1+\frac{\vert y_{L\widetilde{\mathcal{R}}}\vert^2 f^2}{p^2+m_{\widetilde{\mathcal{R}}}^2}\,,\quad \Pi_0^t(p)=1+\frac{\vert y_{R\widetilde{\mathcal{R}}}\vert^2 f^2}{p^2+m_{\widetilde{\mathcal{R}}}^2}\,, \quad \quad \quad \quad  \\ & \Pi_1^q =\frac{\vert y_{L1'}\vert^2 f^2 (m_{\widetilde{\mathcal{R}}}^2-m_{1'}^2)(m_1^2-m_{1^\prime}^2)}{(p^2+m_{\widetilde{\mathcal{R}}}^2)(p^2+m_{1}^2)(p^2+m_{1'}^2)}, \quad \Pi_1^t =\frac{\vert y_{R1'}\vert^2 f^2 (m_{\widetilde{\mathcal{R}}}^2-m_{1'}^2)(m_1^2-m_{1^\prime}^2)}{(p^2+m_{\widetilde{\mathcal{R}}}^2)(p^2+m_{1}^2)(p^2+m_{1'}^2)}\,,\\ &
M_1^t(p)=\frac{y_{L1}y_{R1}^* f^2 m_1}{p^2+m_1^2}-\frac{y_{L\widetilde{\mathcal{R}}}y_{R\widetilde{\mathcal{R}}}^* f^2 m_{\widetilde{\mathcal{R}}}}{p^2+m_{\widetilde{\mathcal{R}}}^2}+\frac{y_{L1'}y_{R1'}^* f^2 m_{1'}}{p^2+m_{1'}^2}\,.
\label{eq: form factors}
\end{aligned}
\end{equation} while, from these sum rules, the relations are derived \begin{equation}
\begin{aligned}
\vert y_{L\widetilde{\mathcal{R}},R\widetilde{\mathcal{R}}}\vert^2 = \vert y_{L1,R1}\vert^2+\vert y_{L1',R1'}\vert^2\,, \quad \vert y_{L\widetilde{\mathcal{R}},R\widetilde{\mathcal{R}}}\vert^2 m_{\widetilde{\mathcal{R}}}^2= \vert y_{L1,R1}\vert^2 m_1^2+\vert y_{L1',R1'}\vert^2 m_{1'}^2\,.
\label{eq: sum rules 2}
\end{aligned}
\end{equation} We have assumed that an extra heavy singlet top-partner, denoted $ T_{1'} $ with mass $m_{1'}$, is included in the above sum rules such that all the integrals converge in the following calculations. Thus, its mass works as a cutoff scale in the calculations. Furthermore, from Eq.~(\ref{eq: sum rules 2}), we can derive the mass hierarchy $m_{1'}>m_{\widetilde{\mathcal{R}}}>m_1$. 

From the effective Lagrangian in Eq.~(\ref{eq: top eff lag}), we can write the Coleman-Weinberg potential in Euclidean momenta space \begin{equation}
\begin{aligned}
V_{\rm top}^{\rm IR}=-2N_c \int \frac{d^4p_E}{(2\pi)^4}\Bigg[&\ln \left(1+\frac{\Pi_1^q}{2\Pi_0^q}\frac{h^2}{\eta^2}\right)+\ln \left(1+\frac{\Pi_1^t}{\Pi_0^t}\left(1-\frac{h^2+\eta^2}{f^2}\right)\right) \\ & + \ln \left(1+\frac{1}{p_E^2} \frac{\vert M_1^t\vert^2}{2\Pi_0^q \Pi_0^t}\frac{h^2}{f^2}\left(1-\frac{h^2+\eta^2}{f^2}\right)\right)\Bigg] \,, \label{eq: pot from top int}
\end{aligned}
\end{equation} where the number of QCD colors is $ N_c=3 $. By expanding the above potential up to the quartic terms of $ h $ and $ \eta $, we can match it to Eq.~(\ref{eq: pot of h eta}), which yields \begin{equation}
\begin{aligned}
&(\mu_h^2)^{\rm IR}_f=-2\alpha_q+\frac{16}{N_L+N_R}\alpha_t-\frac{16}{N_L+N_R} f^2 \beta_t-2f^2 \gamma\,, \\ & (\mu_{\eta}^2)^{\rm IR}_f=\frac{16}{N_L+N_R}\alpha_t - \frac{16}{N_L+N_R}  f^2 \beta_t\,, \quad \quad \quad  \\ & (\lambda_h)^{\rm IR}_f= \beta_q + \frac{64}{(N_L+N_R)^2}\beta_t +4\gamma\,, \quad \quad \quad \\ & (\lambda_\eta)^{\rm IR}_f=\frac{64}{(N_L+N_R)^2}\beta_t\,, \quad  \quad  \quad \\ & (\lambda_{h\eta})^{\rm IR}_f=\frac{64}{(N_L+N_R)^2}\beta_t+2\gamma\,,
\label{eq: IR ferm contribution pot}
\end{aligned}
\end{equation} where {\setlength{\thickmuskip}{.4\thickmuskip}
  \setlength{\medmuskip}{1.1\medmuskip}\begin{equation}
\begin{aligned}
\alpha_{q,t}\equiv \frac{N_c}{f^2} \int \frac{d^4 p_E}{(2\pi)^4}\frac{\Pi_1^{q,t}}{\Pi_0^{q,t}}\,, \quad \beta_{q,t}\equiv \frac{N_c}{f^4}\int \frac{d^4 p_E}{(2\pi)^4}\left(\frac{\Pi_1^{q,t}}{\Pi_0^{q,t}}\right)^2\,, \quad \gamma \equiv \frac{N_c}{f^4}\int \frac{d^4 p_E}{(2\pi)^4}\frac{\vert M_1^t\vert^2}{q^2\Pi_0^q \Pi_0^t}\,.
\label{eq: integrals}
\end{aligned}
\end{equation}}

The potential contributions from the higher order operators arising from top quark loops can be written down using again a spurion analysis~\cite{Alanne:2018wtp} like above for the vector resonances. At leading order in the chiral expansion, we obtain the following potential contributions:
\begin{equation}
\begin{aligned} \label{eq: pot from top int UV 1} V_{\rm top,p^2}^{\rm UV}=&\frac{C_L f^4}{4\pi}y_L^2 \text{Tr}[P_{Q}^\alpha\Sigma^\dagger] \text{Tr}[\Sigma P_{Q\alpha}^{\dagger }]+\frac{C_R f^4}{4\pi}y_R^2 \text{Tr}[P_t\Sigma^\dagger ]\text{Tr}[\Sigma P_t^\dagger]\,,
\end{aligned} \end{equation} 
while a $ y_L^2 $ potential term is not allowed if the symmetric representation is chosen for the left-handed top. At next to the leading order, we obtain \begin{equation}
\begin{aligned} \label{eq: pot from top int UV 2} V_{\rm top,p^4}^{\rm UV}=&\frac{C_{LL} f^4}{(4\pi)^2}y_L^4 \left(\text{Tr}[P_{Q}^\alpha\Sigma^\dagger] \text{Tr}[\Sigma P_{Q\alpha}^{\dagger }]\right)^2+\frac{C_{RR} f^4}{(4\pi)^2}y_R^4 \left(\text{Tr}[P_t\Sigma^\dagger ]\text{Tr}[\Sigma P_t^\dagger]\right)^2\\ &+\frac{C_{LR}f^4}{(4\pi)^2}y_L^2y_R^2\text{Tr}[P_{Q}^\alpha\Sigma^\dagger] \text{Tr}[\Sigma P_{Q\alpha}^{\dagger }]\text{Tr}[P_t\Sigma^\dagger ]\text{Tr}[\Sigma P_t^\dagger]\,.
\end{aligned} \end{equation} Here $C_{L,R}$ and $C_{LL,RR,LR}$ are $\mathcal{O}(1)$ form factors, while $y_{L/R}$ are related to the couplings $\widetilde{y}_{L/R}$ in Eq.~(\ref{eq: top PC-operators}) via the anomalous dimensions of the fermionic operators, and are expected to be $\mathcal{O} (1)$ for the top. The UV contribution from the top quark couplings to the strong sector is thus given by \begin{equation}
\begin{aligned} \label{eq: pot from top int UV} V_{\rm top}^{\rm UV}\equiv V_{\rm top,p^2}^{\rm UV}+V_{\rm top,p^4}^{\rm UV}\,.
\end{aligned} \end{equation} From this effective potential, the contributions to the scalar potential in Eq.~(\ref{eq: pot of h eta}) are then given by \begin{equation}
\begin{aligned}  \label{eq: top parameter contributions}
(\mu_h^2)^{\rm UV}_f=& 4C_L \vert y_L\vert^2 f^2-8C_R \vert y_R\vert^2 f^2 -\frac{2(N_L+N_R)}{\pi^2} C_{RR} \vert y_R\vert^4 f^2 \\ &+\frac{N_L+N_R}{2\pi^2} C_{LR}\vert y_L\vert^2 \vert y_R\vert^2 f^2 \,, \\
 (\mu_\eta^2)^{\rm UV}_f =&-8C_R \vert y_R\vert^2 f^2 -\frac{2(N_L+N_R)}{\pi^2} C_{RR} \vert y_R\vert^4 f^2  \,, \quad \quad \\ 
 (\lambda_h)^{\rm UV}_f=&-\frac{4(N_L+N_R-2)}{N_L+N_R}C_R \vert y_R\vert^2+ \frac{1}{\pi^2}C_{LL}\vert y_L\vert^4 + \frac{6-N_L-N_R}{\pi^2} C_{RR} \vert y_R\vert^4\\ &-\frac{2}{\pi^2}C_{LR} \vert y_L\vert^2 \vert y_R\vert^2\,,
\\ (\lambda_\eta)^{\rm UV}_f=&-\frac{4(N_L+N_R-2)}{N_L+N_R} C_{R} \vert y_R\vert^2 +\frac{6-N_L-N_R}{\pi^2} C_{RR} \vert y_R\vert^4 \,, 
\\ (\lambda_{h \eta})^{\rm UV}_f=& -\frac{4(N_L+N_R-2)}{N_L+N_R}C_R \vert y_R\vert^2 +\frac{6-N_L-N_R}{\pi^2} C_{RR} \vert y_R\vert^4 \\ &-\frac{1}{\pi^2}C_{LR}\vert y_L\vert^2\vert y_R\vert^2\,.
\end{aligned}  
\end{equation}

\subsection{Contributions from the vector-like hyper-fermion masses}

The last potential contributions come from the vector-like masses of the hyper-fermions in Eq.~(\ref{eq: Basic Lagrangian (UV)}), giving rise to \begin{equation}
\begin{aligned}  
V_{\rm m}= & - C_Q f^3 \text{Tr}[M_Q \Sigma_Q]+\rm h.c. \\ = & 4 C_Q m_Q f^3\cos \left(\frac{\pi_Q}{f}\right)\,,  \label{eq: vector-like masses pot}
\end{aligned}
\end{equation} where $ C_Q $ is an $ \mathcal{O}(1) $ non-perturbative coefficient and $ m_Q\equiv \overline{m}_1+\overline{m}_2 $. Note that the vector-like masses $\overline{m}_i$ with $i\geq 3$ have no influence on the following calculations. According to this effective potential, the contributions to the scalar potential in Eq.~(\ref{eq: pot of h eta}) are then \begin{equation}
\begin{aligned} 
(\mu_h^2)_m=& (\mu_\eta^2)_m= -8C_Q m_Q f\,, \quad \quad 
\\ (\lambda_h)_m=&(\lambda_\eta)_m=(\lambda_{h \eta})_m=\frac{4C_Q m_Q}{3f}\,.
\end{aligned}  
\end{equation}

\section{Numerical results for the phase transition and GW generation}
\label{sec: Numerical results}

In this section, we perform a numerical analysis of the gravitational wave signals generated during SFOEWPTs for the CH models with cosets $ \SU(N)/\SP(N) $. The numerical
calculations are based on a finite temperature scalar potential to all orders by using the $\textsf{CosmoTransitions} $ package~\cite{Wainwright:2011kj}, and not only to the leading order of $ T^2 $ as in Eq.~(\ref{eq: finite temperature potential}), but we note that the calculations from the full finite temperature potential or the leading order potential give rise to almost indistinguishable results. 

\subsection{The model parameters}

There are five free parameters in the finite temperature potential in Eq.~(\ref{eq: finite temperature potential}): $ \mu_h^2 $, $ \mu_\eta^2 $, $ \lambda_h $, $ \lambda_\eta $ and $ \lambda_{h\eta} $. For fixed PNGB decay constant $ f $ in the Higgs mass expression in Eq.~(\ref{eq: Higgs and eta mass expressions}), we can fix $ \mu_h^2 $, while the quartic coupling $ \lambda_h $ is fixed by the EW VEV expression in Eq.~(\ref{eq: VEV expressions}). Finally, we can replace the mass parameter $ \mu_\eta^2 $ with the $ \eta $ mass, $ m_\eta $, via Eq.~(\ref{eq: Higgs and eta mass expressions}). Thus, for a fixed value of $ f $, we have three free parameters to scan, $ m_\eta $, $ \lambda_{\eta}$ and $ \lambda_{h\eta} $, to find degenerate
vacua that fulfill the conditions in Eqs.~(\ref{eq: degenerate vacuums condition}),~(\ref{eq: critical condition}) and~(\ref{eq: strong criterion}), leading to SFOEWPTs.

Furthermore, the free parameters in a CH model with the coset $ \SU(N)/\SP(N) $ are as follows\begin{equation}
\begin{aligned} 
f\,, \, m_V\,,\, m_A\,, \, m_1 \,, \, m_{\widetilde{\mathcal{R}}}\,, \, m_{1'}\,, \, y_{L\widetilde{\mathcal{R}}}\,, \, y_{R\widetilde{\mathcal{R}}}\,, \, m_Q \,,
\end{aligned}  
\end{equation} where $ f_{K,1} $ and $ y_{L1,L1',R1,R1'} $ are fixed by the sum rules in Eqs.~(\ref{eq: sum rule gauge 2}) and~(\ref{eq: sum rules 2}). We scan over these parameters as follows: $ 2~\text{TeV}\leq m_{V,A}\leq 7~\text{TeV}$,  $1~\text{TeV} \leq  m_{1,\widetilde{\mathcal{R}},1'} \leq 6~\text{TeV} $, $ \vert y_{i} \vert \leq 5 $, $ 0.01 f\leq m_Q \leq f $, while we consider two different values of the PNGB decay constant $ f=1.0, \,2.5 $~TeV. Moreover, we require that the absolute values of the non-perturbative coefficients $ c_{g,g'} $, $ d_{g,g'} $, $ C_{L,R} $, $ C_{LL,RR,LR} $ and $ C_Q $ are smaller than $ 5 $. 

\begin{figure}[t]
	\centering
\includegraphics[width=0.95\textwidth]{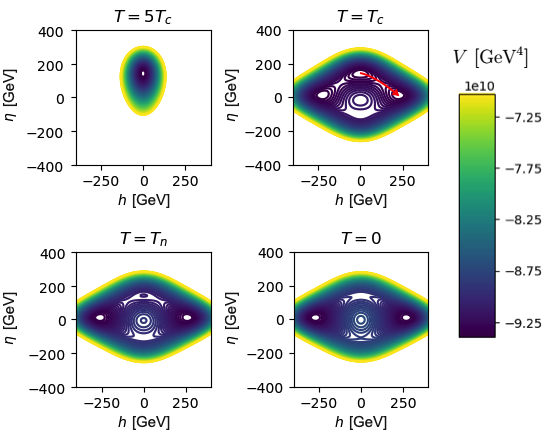}
	\caption{The finite temperature potential of $ h $ and $ \eta $ before the phase transition at $ T=5T_{c} $ (top left), and close to the first-order phase transition at the critical and nucleation temperatures $ T_{c} $ (top right) and $ T_{n} $ (bottom left), and finally at $ T=0 $ (bottom right). The red arrow depicts the movement of the ground state under the phase transition. }
	\label{fig: phase transitions}
\end{figure}

\subsection{SFOEWPTs from the CH dynamics}

We observe first-order phase transitions in parameter space as shown in Figure~\ref{fig: phase transitions}, giving rise to a strong signal of gravitational waves. The hierarchy of the temperatures in the various panels in the figure is $ 5T_{c}>T_{c}>T_{n}>0$ (decreasing from upper left to lower right panel), while the red arrows depict the movement of the ground state in the potential under the phase transition. 

Before the phase transition at the temperature $ T=5T_{c} $, there is only one minimum in the potential leading to $ \langle h \rangle =0 $ and $ \langle \eta \rangle \neq 0 $ as expected. 
During the phase transition, about $ T=T_{c} $, the state of the vacuum changes from zero VEV in the direction of $ h $ to a non-zero VEV, while the $ \eta $ field decreases to zero. As we will see, these SFOEWPTs can lead to significant gravitational wave signals. Finally, at zero temperature, the barriers between the four minima and the maximum in their center have been increased sufficiently such that the vacuum state is locked to the minimum at the right with a positive non-zero VEV of $ h $ ($\langle h \rangle = v_{\rm SM}$). However, if we perform the same analysis with the leading order potential given in Eq.~(\ref{eq: finite temperature potential}), this picture changes insignificantly.

\begin{figure}[t]
	\centering
	\includegraphics[width=0.72\textwidth]{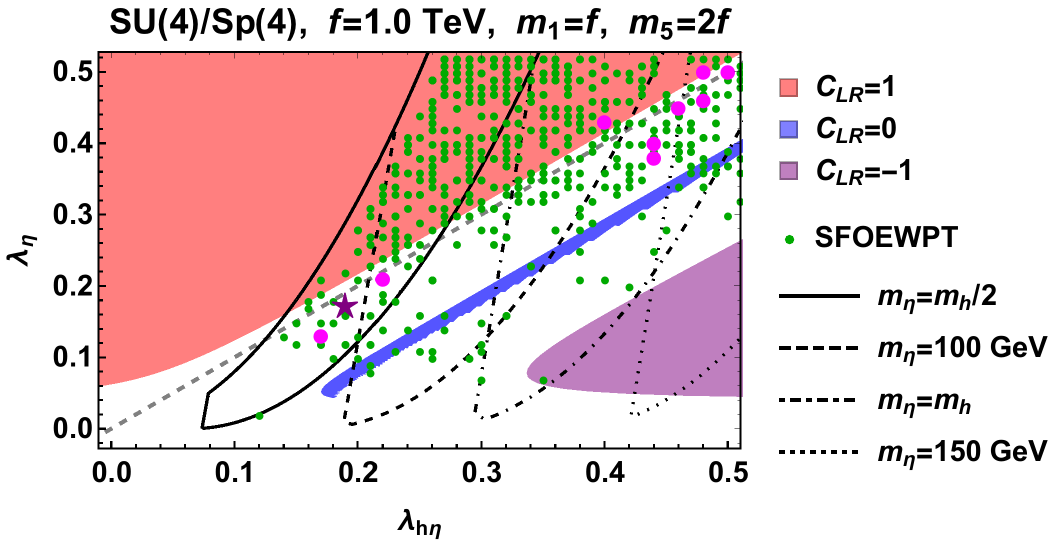}
	\includegraphics[width=0.72\textwidth]{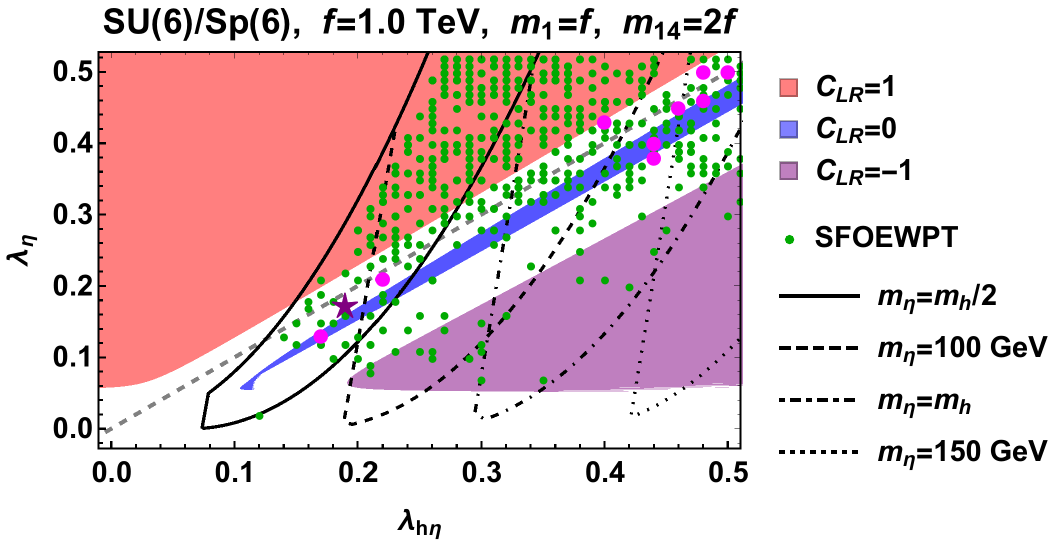}
	\includegraphics[width=0.72\textwidth]{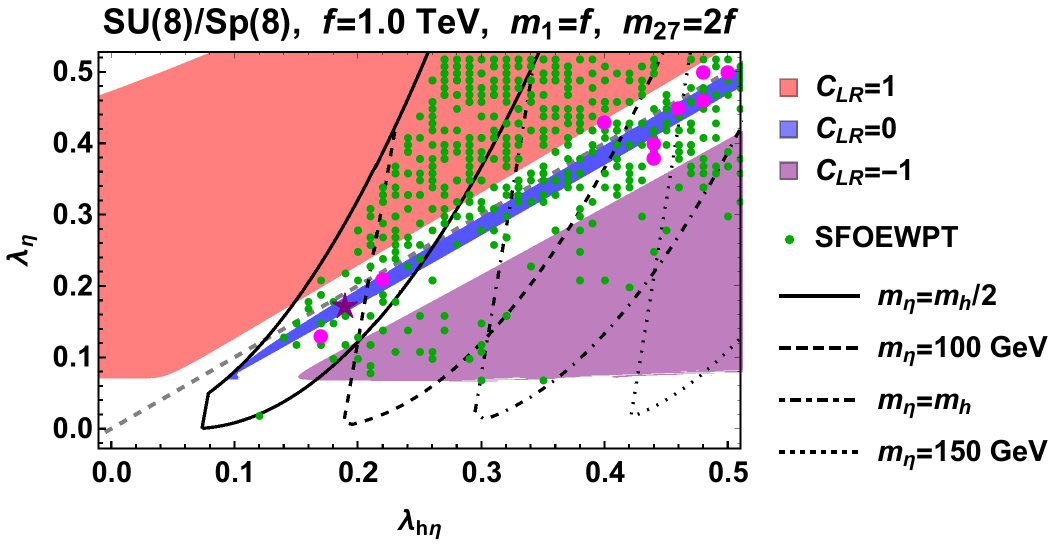}
	\caption{Allowed regions in the $ \lambda_{h\eta} $--$ \lambda_\eta $ parameter space for $ N_L+N_R=2,3,4 $ (i.e. $ \SU(4)/\SP(4) $, $ \SU(6)/\SP(6) $ and $ \SU(8)/\SP(8) $ CH models) for various $ C_{LR} =0,1,-1 $. We have fixed $ f=1.0 $~TeV and $ m_1=m_{\widetilde{\mathcal{R}}}/2=f $, while we scan over $ m_{1^\prime}=2f\dots 6f $ and all the mixing couplings $ \vert y_{L,R}^{1,\widetilde{\mathcal{R}},1^\prime}\vert $ within 5. The black lines delimit the areas providing $ m_\eta=m_h/2,100~\text{GeV},m_h,150 $~GeV. The gray line represents $ \lambda_\eta=\lambda_{h\eta} $. The green points can trigger SFOEWPTs. The parameter points marked with purple stars lead to GWs that may be measured by the planned LISA space probe, depicted in the upper panel in Figure~\ref{fig: The GW signals} by the three blue curves exceeding the LISA sensitivity (the purple line). The magenta points along with the stars represent the ten GW strength curves in Figure~\ref{fig: The GW signals} with the largest peak amplitudes. }
	\label{fig: figure 1}
\end{figure}

\begin{figure}[t]
	\centering
	\includegraphics[width=0.72\textwidth]{fig/SU4_1_1.pdf}
	\includegraphics[width=0.72\textwidth]{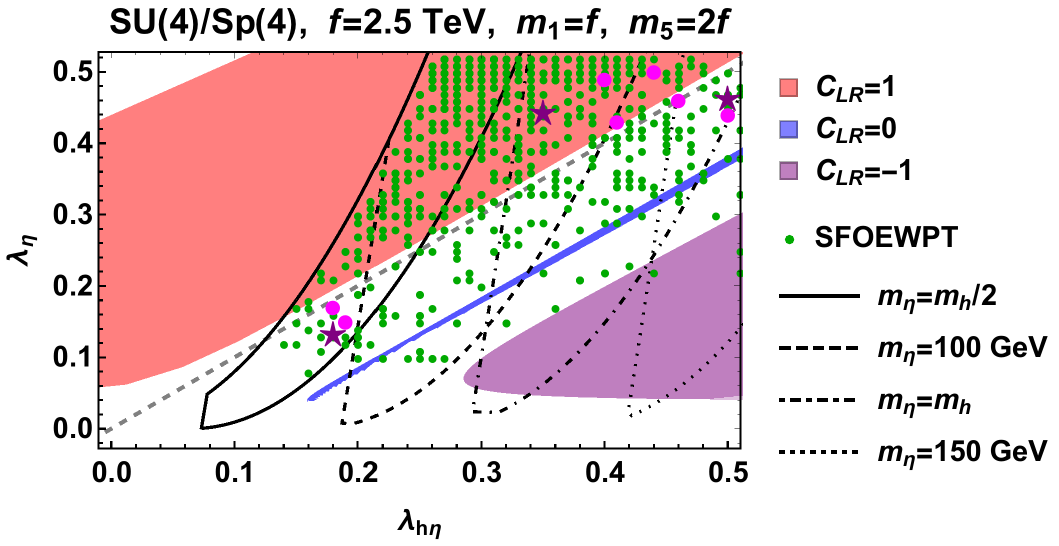}
	\caption{Allowed regions in the $ \lambda_{h\eta} $--$ \lambda_\eta $ parameter space for the minimal $ \SU(4)/\SP(4) $ CH model (i.e. $ N_L+N_R=2 $) for various $ C_{LR} =0,1,-1 $. We have fixed $ f=1.0 $~TeV in the upper panel and $ f=2.5 $~TeV in the lower panel, while $ m_1=m_{\widetilde{\mathcal{R}}}/2=f $. We scan over $ m_{1^\prime}=2f\dots 6f $ and all the mixing couplings $ \vert y_{L,R}^{1,\widetilde{\mathcal{R}},1^\prime}\vert $ within 5. The black lines delimit the areas providing $ m_\eta=m_h/2,100~\text{GeV},m_h,150 $~GeV. The gray line represents $ \lambda_\eta=\lambda_{h\eta} $. The green points can trigger SFOEWPTs. Finally, the parameter points marked with purple stars lead to GWs that may be measured by the planned LISA space probe, depicted in Figure~\ref{fig: The GW signals} by the blue and red curves exceeding the LISA sensitivity (the purple line) for $ f=1.0 $~TeV and $ f=2.5 $~TeV, respectively. The magenta points along with the stars represent the ten GW strength curves in Figure~\ref{fig: The GW signals} with the largest peak amplitudes.  }
	\label{fig: figure 2}
\end{figure}

\begin{figure}[t]
	\centering
	\includegraphics[width=0.72\textwidth]{fig/SU4_1_1.pdf}
	\includegraphics[width=0.72\textwidth]{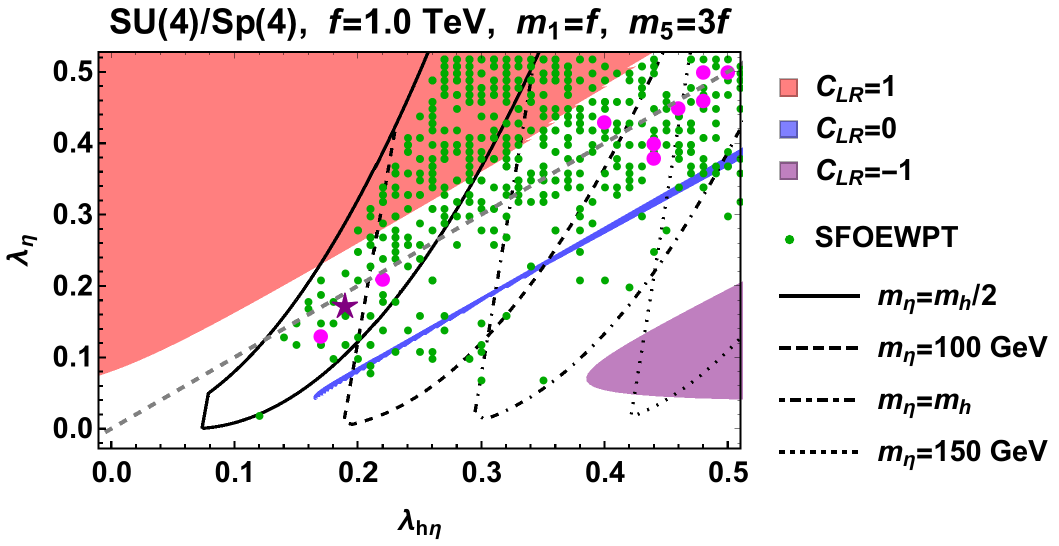}
	\includegraphics[width=0.72\textwidth]{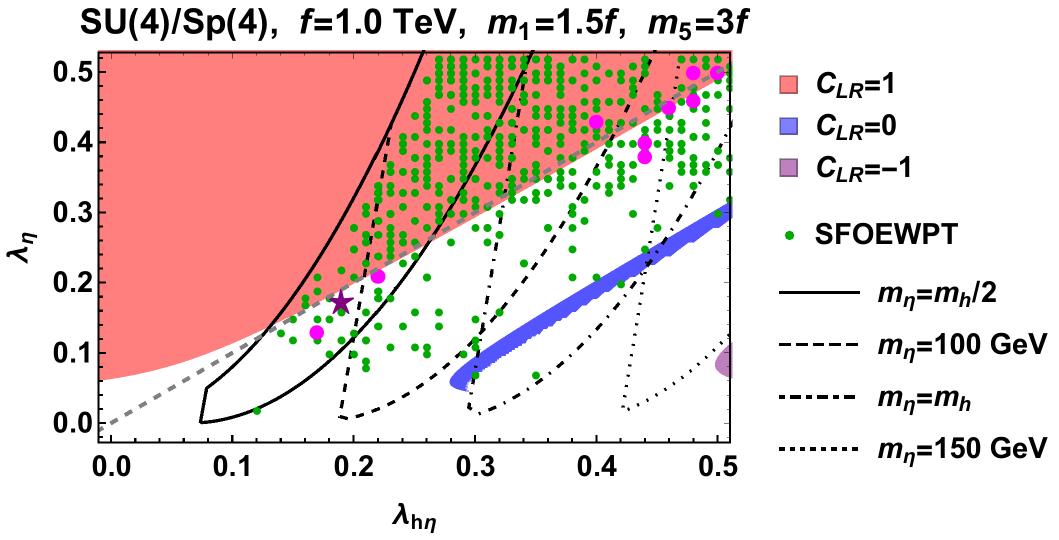}
	\caption{Allowed regions in the $ \lambda_{h\eta} $--$ \lambda_\eta $ parameter space for varying mass parameters of the fermion resonances with $ N_L+N_R=2 $ (i.e. the minimal $ \SU(4)/\SP(4) $ CH model). We have fixed $ f=1.0 $~TeV for various $ C_{LR} =0,1,-1 $, while we scan over $ m_{1^\prime}=m_5\dots 6f $ and all the mixing couplings $ \vert y_{L,R}^{1,\widetilde{\mathcal{R}},1^\prime}\vert $ within 5. The black lines delimit the areas providing $ m_\eta=m_h/2,100~\text{GeV},m_h,150 $~GeV. The gray line represents $ \lambda_\eta=\lambda_{h\eta} $. The green points can trigger SFOEWPTs. Finally, the parameter points marked with purple stars lead to GWs that may be measured by the planned LISA space probe, depicted in the upper panel in Figure~\ref{fig: The GW signals} by the three blue curves exceeding the LISA sensitivity (the purple line). The magenta points along with the stars represent the ten GW strength curves in Figure~\ref{fig: The GW signals} with the largest peak amplitudes. }
	\label{fig: figure 3}
\end{figure}

In Figures~\ref{fig: figure 1},~\ref{fig: figure 2} and~\ref{fig: figure 3}, the regions delimited by the black lines represent the $ \lambda_{h\eta} $--$ \lambda_\eta $ parameter space for various $ m_\eta $ providing degenerate vacua by fulfilling the conditions in Eq.~(\ref{eq: degenerate vacuums condition}). The decay process of the Higgs into a pair of $ \eta $ states leads to a constraint on the $ \eta $ mass by experimental measurements~\cite{ATLAS:2018doi,CMS:2018uag}. Due to the fact that the Higgs total width is very small $ \Gamma_h=4.07 $~MeV~\cite{Denner:2011mq} and the coupling $ \lambda_{h\eta} $ needs to be sizable according to Eq.~(\ref{eq: degenerate vacuums condition}), the decay channel $ h\rightarrow \eta\eta $ would dominate the Higgs decay and we will, therefore, assume $ m_\eta > m_h/2 $ to meet the above constraints. Moreover, in these figures, the colored shaded areas represent the regions in the $ \lambda_{h\eta} $--$ \lambda_\eta $ parameter space that are spanned by scanning over the parameters of various $ \SU(N)/\SP(N) $ CH models. Of the various parameters, these shaded areas are most sensitive to the value of the non-perturbative coefficient $ C_{LR} $ given in the potential term that mixes the left- and right-handed top spurions in Eq.~(\ref{eq: pot from top int UV 2}), where the purple, blue and red areas are associated with the values $ C_{LR}=-1,0,1 $, respectively. 
In between the outer boundaries of the red and purple regions, the coefficient $ C_{LR} $ can take different values in the interval $ -1\leq C_{LR}\leq 1 $. Additionally, all the results in these figures are independent of the number of SM singlet hyper-fermions, $ N_\Lambda $, and we thus ignore $ N_\Lambda $ in the following. Furthermore, the green points can trigger SFOEWPTs, since they fulfill the conditions in both Eq.~(\ref{eq: degenerate vacuums condition}), the critical condition in Eq.~(\ref{eq: critical condition}) (providing first-order phase transitions) and the criterion in Eq.~(\ref{eq: strong criterion}) (ensuring that they are strong). If these parameter points lie in the colored shaded areas, the CH models can give rise to these SFOEWPTs. Moreover, the parameter points marked with purple stars lead to GWs that may be measured by the planned LISA space probe, depicted in Figure~\ref{fig: The GW signals} (discussed in Section~\ref{sec: Gravitational waves generated in the CH models}) by the curves exceeding the LISA sensitivity (the purple line). Finally, the magenta points along with the purple stars represent the ten GW strength curves in Figure~\ref{fig: The GW signals} with the largest peak amplitudes among all the green parameter points.

In Figure~\ref{fig: figure 1}, the allowed regions are shown in the $ \lambda_{h\eta} $--$ \lambda_\eta $ parameter space for $ N_L + N_R = 2,3,4 $, i.e. the cosets $ \SU(4)/\SP(4) $ (upper), $ \SU(6)/\SP(6) $ (middle) and $ \SU(8)/\SP(8) $ (lower panel), respectively. In this figure, we have fixed $ f = 1 $~TeV and $ m_1 = m_{\widetilde{\mathcal{R}}}/2 = f  $ in order to compare the various CH models.
In Figures~\ref{fig: figure 2} and~\ref{fig: figure 3}, we will also investigate the influence on our results by changing $ f $ and the masses $ m_1 , m_{\widetilde{\mathcal{R}}} $, respectively. These results are in agreement with the analysis in Ref.~\cite{Bian:2019kmg}, where they study the generation of GWs from the $ \SO(6)/\SO(5)\sim \SU(4)/\SP(4) $ CH model. This model is similar to the $ \SU(4)/\SP(4) $ CH model considered here except that the global symmetry $ \SU(4) $ is explicitly broken to $ \SP(4) $ by adding vector-like masses of the hyper-fermions. In contrast, the $ \SO(6)/\SO(5) $ coset is thus difficult to realize in a simple manner from a four-dimensional gauge-fermion theory, but it arises more simply in the five-dimensional holographic approach~\cite{Agashe:2004rs}. Furthermore, in their analysis of the $ \SO(6)/\SO(5) $ coset, they have not included the allowed $ C_{LR} $ potential term in Eq.~(\ref{eq: pot from top int UV 2}). Therefore, their analysis covers only the narrow blue shaded areas (i.e. $ C_{LR}=0 $) shown in the figures, resulting in an incomplete analysis compared to ours. According to the figures, when this term ($ C_{LR}\neq 0 $) is turned on, the parameter space opens up significantly. For $ C_{LR}\leq 0 $, the allowed area is always located below the gray dashed line in the figures, i.e. $ \lambda_\eta < \lambda_{h\eta} $. In this region of the parameter space, the frequency of SFOEWPT points (the green points) is significantly smaller compared to the region above the line, i.e. $ \lambda_\eta> \lambda_{h\eta} $. On the other hand, the frequency of points leading to SFOEWPTs and hence GWs is much larger for $ C_{LR} > 0 $.
This is illustrated by the example with $ C_{LR}=1 $ (the red shaded area) in the figures, which covers almost the entire parameter space above the gray line. However, as illustrated in Figure~\ref{fig: figure 1}, the allowed area in the parameter space with $ C_{LR}\lesssim 0 $ (e.g. the blue and purple areas) will cover increasing more SFOEWPT points as $ N_L+N_R $ increases due to the fact that these areas move up towards where the concentration of SFOEWPT points is larger. Furthermore, to obtain SFOEWPTs that may be measurable by the planned LISA space probe (marked with stars), we need that the non-perturbative coefficient $ C_{LR} $ takes a positive value somewhere in between $ 0<C_{LR}<1 $ for $ \SU(4)/\SP(4) $ and $ \SU(6)/\SP(6) $, while $ C_{LR}=0 $ may also be possible for $ \SU(8)/\SP(8) $. Finally, these star points together with the magenta points, providing the largest peak amplitudes of the GW strength curves shown in Figure~\ref{fig: The GW signals}, indicate that parameter points with large peak amplitudes lie around the gray line (i.e. $ \lambda_\eta=\lambda_{h\eta} $).

In Figure~\ref{fig: figure 2}, we show the allowed regions in the $ \lambda_{h\eta} $--$ \lambda_\eta $ parameter space for the minimal CH model with the coset $ \SU(4)/\SP(4) $ (i.e. $ N_L+N_R =2 $), where we consider the cases with $ f=1.0$~TeV ($ s_\theta \approx 0.25 $) and $ f=2.5$~TeV ($ s_\theta \approx 0.10 $) in the upper and lower panel, respectively. The reason that we consider the first benchmark point with   $ f=1.0$~TeV ($ s_\theta \approx 0.25 $) is that it satisfies the constraint by direct LHC measurements~\cite{deBlas:2018tjm} of the CH Higgs coupling to the EW gauge bosons and top quark, implying an upper bound of $ f\gtrsim 0.8 $~TeV ($ s_\theta \lesssim 0.3 $) discussed below Eq.~(\ref{SM VEV}). The second benchmark point with $ f=2.5$~TeV ($ s_\theta \approx 0.10 $) is chosen such that the vacuum misalignment angle, $ \theta $, is maximally fine-tuned by one part in 100, minimizing the so-called little hierarchy problem of the CH framework.~\footnote{The little hierarchy problem in the CH framework is typically quantified in terms of the fine-tuning parameter $ \xi\equiv (v_{\rm SM}/f)^2 = 1/s_\theta^2 $. For example, when $ s_\theta=0.1 $, the misalignment angle in this CH model is fine-tuned by one part in 100. A smaller $ s_\theta $, therefore, requires more unnatural fine-tuning.} From Figure~\ref{fig: figure 2}, we can conclude that the blue shaded area (with $ C_{LR}=0 $) shrinks when $ f $ increases, leading to a smaller parameter space triggering SFOEWPTs. This is again in agreement with the results in Ref.~\cite{Bian:2019kmg} for the $ \SO(6)/\SO(5) $ CH model, where they show in the left panel in Figure~2 that the frequency of SFOEWPTs decreases as $ f $ increases. Moreover, we note that the allowed regions for $ C_{LR}=\pm 1 $ (the red and purple areas) and the distribution of the SFOEWPT points change slightly when $ f $ is varied. Also for a larger $ f $, the parameter points (the purple stars and the magenta points), providing the largest peak amplitudes of the GW strength curves shown in Figure~\ref{fig: The GW signals}, are distributed around the gray line (i.e. $ \lambda_\eta=\lambda_{h\eta} $), favoring scenarios with positive $C_{LR}$. However, there is a greater spread of them for larger $ f $. 

In Ref.~\cite{DeCurtis:2019rxl}, they investigate the generation of GWs within the context of the SO(6)/SO(5) CH model, considering specific $\eta$ masses of 150 GeV and 250 GeV. Our findings are consistent with the outcomes presented in the left panel of Figure 3 from the aforementioned reference. In our study, we have chosen the (6,6) representation for both left- and right-handed top quark spurions (the two-index antisymmetric representation of SU(4)), which is the minimal choice. For the model considered in Ref.~\cite{DeCurtis:2019rxl}, this corresponds to the fundamental representation of SO(6), i.e. $(6,6)$. In their Figure~2, they compare the parameter space covered by various CH models, where the dashed lines depict the maximal values allowed for the quartic coupling $ \lambda_{h\eta} $. For the $ (6,6) $ example, the maximal value is $\lambda_{h\eta}< \lambda_h \approx 0.13 $. However, due to the fact that the coefficient $C_{LL}$ of the $y_L^4$ term in Eq.~(\ref{eq: pot from top int UV 2}) is negative to achieve the correct vacuum misalignment, it negatively contributes in Eq.~(\ref{eq: top parameter contributions}) to $ \lambda_h $, while $ \lambda_{h\eta} $ remains unaffected, removing the inequality. This is also supported by Ref.~\cite{Bian:2019kmg}.

Finally, in Figure~\ref{fig: figure 3}, we vary the mass parameters of the fermion resonances ($ m_1=f $ and $ m_5=2f $ for upper panel, $ m_1=f $ and $ m_5=3f $ for middle panel, and $ m_1=1.5 f $ and $ m_5=3f $ for lower panel) for the minimal coset $ \SU(4)/\SP(4) $ (i.e. $N_L+N_R =2$) and $ f=1.0 $~TeV. From this figure, we conclude that when we increase the mass parameters of the fermion resonances the regions with $ C_{LR}\leq 0 $ are reduced since these regions either move away from the SFOEWPT points or become more restricted. In addition, the red region with $ C_{LR}= 1 $ will be reduced slightly by increasing the difference between $m_1$ and $ m_5 $ (illustrated in the middle panel) opening up the parameter space with $ 0<C_{LR}<1 $, while if we keep the ratio $m_5/m_1=2$ for various mass parameters (the upper and lower panels) this region is almost unchanged.

Conclusively, according to the analysis in this section, a CH model with a positive $ C_{LR} $ is more testable by GWs due to the large frequency of points triggering SFOEWPTs for $ C_{LR}>0 $. Moreover, positive values of $ C_{LR} $ are further favored when $ f $ and $ m_{1,\widetilde{R}} $ are increased since the allowed regions with $ C_{LR} \leq 0 $ are reduced in at least one of these cases. However, the parameter space with $ C_{LR} \leq 0 $ can be opened up by extending the CH coset as shown in Figure~\ref{fig: figure 1}. Finally, the CH models with the parameters generating measurable GWs by LISA (marked by the purple stars) require a positive $ C_{LR} $ except if the coset of the CH model is minimally extended to a $ \SU(8)/\SP(8) $ coset.

\begin{figure}[t]
	\centering
	\includegraphics[width=0.50\textwidth]{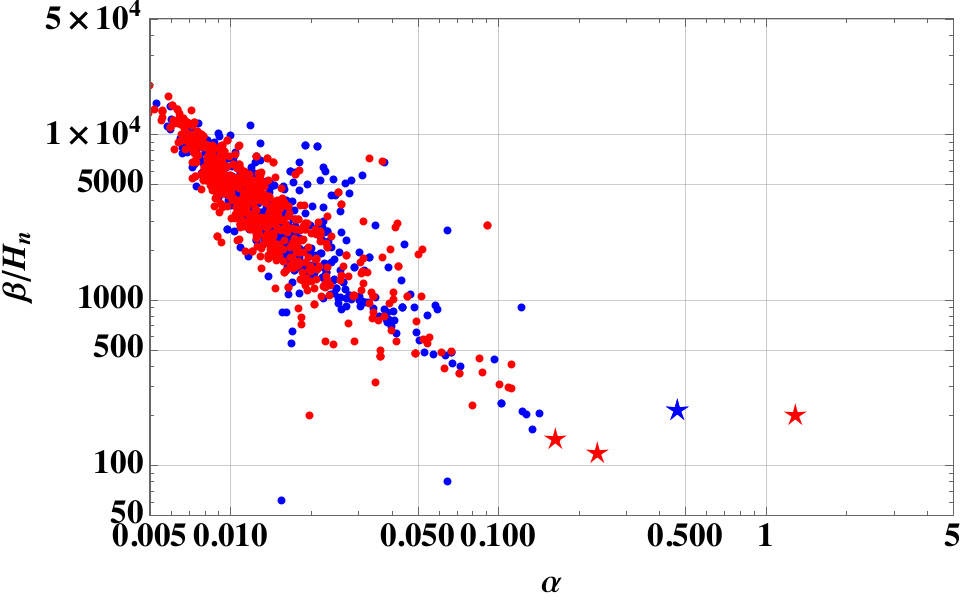}
		\includegraphics[width=0.48\textwidth]{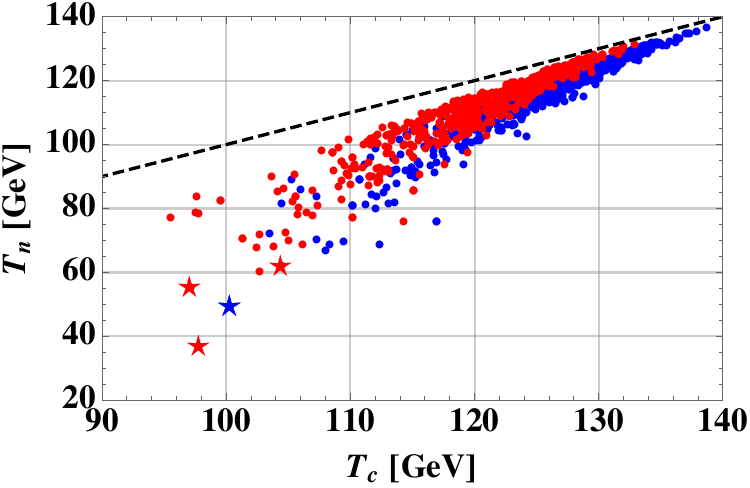}
	\caption{Left panel: points triggering SFOEWPTs in the $ \alpha $--$ \beta/H_n $ parameter space. Right panel: the parameter points from the left
panel in the $ T_c $--$ T_{n} $ parameter space. For the blue (red) parameter points, the decay constant is fixed $ f=1.0 $~TeV ($ 2.5 $~TeV). The parameter points marked with stars lead to GWs that may be measured by the planned LISA space probe, depicted in Figure~\ref{fig: The GW signals} by the curves exceeding the LISA sensitivity (the purple line). }
	\label{fig: alpha beta Tc Tn}
\end{figure}

\subsection{Gravitational waves generated in the CH models}
\label{sec: Gravitational waves generated in the CH models}

As observed until now, these CH models undergo a phase transition, where the new strong dynamics confines at about the temperature $ T_{\rm HC}\sim f $. However, according to these CH models, the new strong dynamics may also lead to a ``two-step'' phase transition, where the VEVs $ (\langle h \rangle, \langle \eta \rangle) $ of the composite fields $ h $ and $ \eta $ develop as $ (0,0)\rightarrow (0,v_\eta) \rightarrow (v_{h},0) $ when the universe cooled down from $ T\gg m_h $ to $ T\approx 0 $. If these two vacua are degenerated at some critical temperature $ T_c $ by fulfilling the conditions in Eq.~(\ref{eq: degenerate vacuums condition}) and the critical condition Eq.~(\ref{eq: critical condition}), this phase transition will be identified as first-order, and if the strong criterion in Eq.~(\ref{eq: strong criterion}) is also satisfied a SFOEWPT is achieved. Therefore, there should exist a nucleation temperature $ T_n $ fulfilling $ S_3(T_n)/T_n\sim 140 $. 
\begin{figure}[t]
	\centering
		\includegraphics[width=0.89\textwidth]{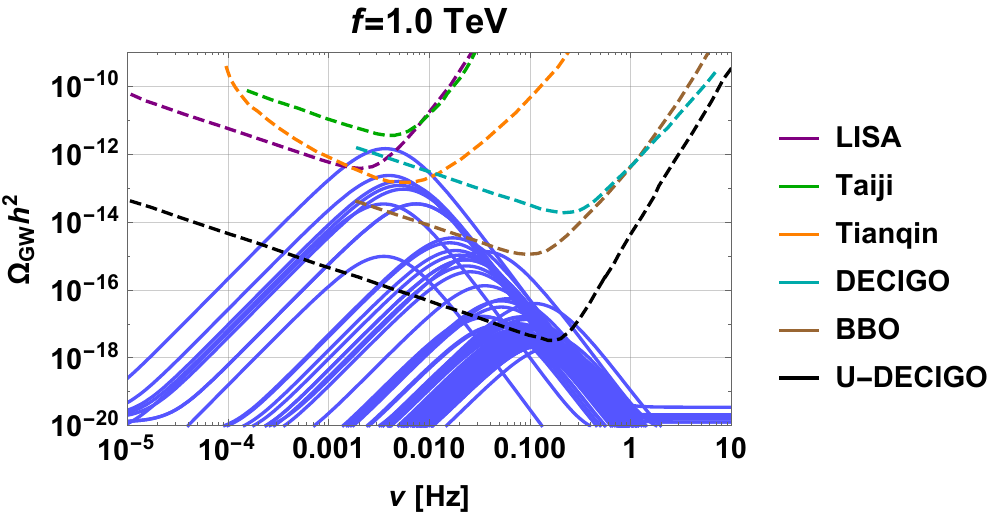} \\ \tiny \quad \\
			\includegraphics[width=0.89\textwidth]{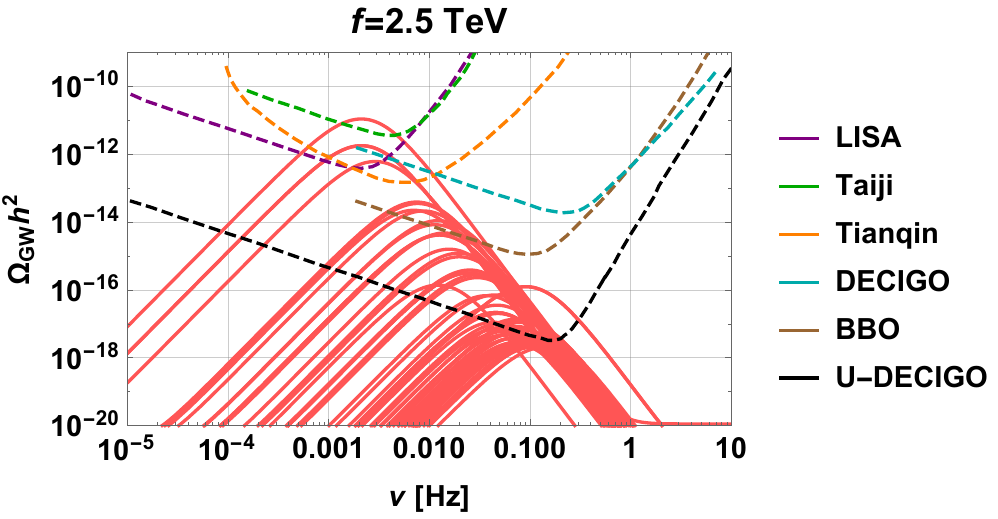}
	\caption{The GW signals, where the solid lines are the GW curves from the parameter points in Figure~\ref{fig: alpha beta Tc Tn} while the dashed lines represent the sensitivities for various future GW detectors such as LISA~\cite{LISA:2017pwj}, Taiji~\cite{Hu:2017mde}, Tianqin~\cite{TianQin:2015yph}, BBO~\cite{Crowder:2005nr}, DECIGO~\cite{Kawamura:2006up,Kawamura:2011zz} and Ultimate-DECIGO (U-DECIGO)~\cite{Kawamura:2011zz}. The blue (red) lines in the upper (lower) panel represent the GW curves from the parameter points with $ f=1.0 $~TeV ($ 2.5 $~TeV).}
	\label{fig: The GW signals}
\end{figure}
Such a SFOEWPT may generate GWs~\cite{Grojean:2006bp}. During a first-order phase transition (FOPT), GWs can be generated by bubbles collisions~\cite{Huber:2008hg,Jinno:2016vai}, stirred acoustic waves~\cite{Hindmarsh:2015qta,Hindmarsh:2017gnf} and magnetohydrodynamic turbulence in the super-cooling plasma~\cite{Caprini:2009yp,Binetruy:2012ze}. The GW spectrum is \begin{equation} \begin{aligned} \label{eq: total GW power spectrum}
\Omega_{\rm GW}(\nu)=\frac{1}{\rho_c}\frac{d\rho_{\rm GW}}{d\ln \nu}\,,
\end{aligned}\end{equation} where $ \rho_c = 3H_0^2 /(8\pi G) $ is the critical energy density today, $ \nu $ denotes the frequency of the GWs and $\rho_{\rm GW} $ is the GW energy density produced during FOPT. Here, $H_0 = 100 h~\text{km~s}^{-1}\text{Mpc}^{-1} $ is the Hubble rate today with $h=0.72$. 

The GW spectrum from FOPTs is generally characterized by three essential parameters, $ \alpha $, $ \beta/H_n $ and $ v_w $, evaluated at the nucleation temperature, $ T_n $. The parameter $ \alpha $ is the ratio of the vacuum energy density $ \epsilon $ and radiation energy density $ \rho_{\rm rad} $, while $ \beta /H_n $ is the nucleation rate divided with the Hubble rate evaluated at $ T_n $ which measures the time duration of the phase transition. Thus, the smaller $ \beta /H_n $ is, the stronger the phase transition is. Finally, $ v_w $ is the wall velocity of the FOPT bubble, which is calculated in the following by a simple approximation given by Ref.~\cite{Lewicki:2021pgr}. These parameters are, therefore, given by \begin{equation} \begin{aligned}\label{eq: alpha beta/Hn}
\alpha \equiv \frac{\epsilon}{\rho_{\rm rad}}\,,   \quad\quad \frac{\beta}{H_n}\equiv T_n \frac{d}{dT}\left(\frac{S_3}{T}\right)\bigg \vert_{T=T_n}\,, \quad v_w =\Bigg\{\begin{array}{cl}
\sqrt{\frac{\Delta V_T}{\epsilon}} & \quad\text{for} \quad \sqrt{\frac{\Delta V_T}{\epsilon}}< v_J(\alpha)  \\
1 &\quad \text{for} \quad \sqrt{\frac{\Delta V_T}{\epsilon}} \geq  v_J(\alpha)  \end{array}
\end{aligned}\end{equation} where \begin{equation} \begin{aligned}
\epsilon=&-\Delta V_T +T_n\Delta\frac{\partial V_T}{\partial T} \bigg\vert_{T=T_n}\,, \quad\quad \rho_{\rm rad}= \frac{\pi^2}{30}g_* T_n^4\,, \quad\quad \beta=\frac{d}{dt}\left(\frac{S_3}{T}\right)\bigg \vert_{t=t_n}\,,
\end{aligned}\end{equation} and \begin{equation} \begin{aligned}
v_J(\alpha)=&\frac{1}{\sqrt{3}}\frac{1+\sqrt{3\alpha^2+2\alpha}}{1+\alpha}\,.
\end{aligned}\end{equation} Here ``$ \Delta $'' denotes the difference between the true and false vacua and $v_J$ is the Jouguet velocity~\cite{Steinhardt:1981ct,Kamionkowski:1993fg,Espinosa:2010hh}, while $ g_* $ and $ t_n $ are the relativistic degrees of freedom and the cosmic time at $ T_n $, respectively.

The total power spectrum of the GWs in Eq.~(\ref{eq: total GW power spectrum}) consists of the three above-mentioned components, which can be written as \begin{equation} \begin{aligned}\label{eq: total power spectrum}
\Omega_{\rm GW}(\nu)h^2=\Omega_{\rm env}(\nu)h^2+\Omega_{\rm sw}(\nu)h^2+\Omega_{\rm turb}(\nu)h^2\,, 
\end{aligned}\end{equation} where $ \Omega_{\rm env}(\nu)h^2 $, $ \Omega_{\rm sw}(\nu)h^2 $ and $ \Omega_{\rm turb}(\nu)h^2 $ are the power spectra of the GWs, respectively, from bubble collisions, sound waves and turbulence of the magnetohydrodynamics in the particle bath. The full expressions of these spectra are given in Refs.~\cite{Ellis:2018mja, Ellis:2020awk}. These expressions are modified compared to them given by e.g. Ref.~\cite{Chen:2017cyc}, since the sound wave period does not last a full Hubble time. This results in a reduction of the total GW signal. In Refs.~\cite{Ellis:2019oqb, Ellis:2020nnr}, it is further shown that bubble collisions are
almost negligible in transitions with polynomial potentials. These issues are also reviewed by LISA in Ref.~\cite{Caprini:2019egz}. Therefore, we will in the following neglect these contributions, i.e. $\Omega_{\rm GW}(\nu)\approx \Omega_{\rm sw}(\nu)+\Omega_{\rm turb}(\nu) $. According to Ref.~\cite{Ellis:2020awk}, the sound wave contribution can be written as \begin{equation} \begin{aligned}\label{eq: sw GW power spectrum}
\Omega_{\rm sw}(\nu)h^2 \simeq& 4.18\times 10^{-6} v_w (H_n\tau_{\rm sw}) \left(\frac{100}{g_*}\right)^{1/3} \left(\frac{\beta}{H_n}\right)^{-1} \left(\frac{\kappa_{\rm v}\alpha}{1+\alpha}\right)^2 \left(\frac{\nu}{\nu_{\rm sw}}\right)^3 \\&\times \left[1+\frac{3}{4}\left(\frac{\nu}{\nu_{\rm sw}}\right)^2\right]^{-7/2}\,,
\end{aligned}\end{equation} where the peak frequency is given by \begin{equation} \begin{aligned}\label{eq: sw peak frequency}
\nu_{\rm sw}=1.91\times 10^{-5}~\text{Hz}\left(\frac{T_n}{100~\text{GeV}}\right)(1+\alpha)^{1/4}\left(\frac{g_*}{100}\right)^{1/6}\frac{1}{v_w}\left(\frac{\beta}{H_n}\right)\,,
\end{aligned}\end{equation} while the duration of the sound wave source can be approximated as follows~\cite{Ellis:2020awk} \begin{equation} \begin{aligned}\label{eq: sound wave period }
H_n \tau_{\rm sw}\approx \text{min}\left[1, (8\pi)^{1/3}\text{max}(v_w,c_s) \sqrt{\frac{4}{3}\frac{1+\alpha}{\kappa_{\rm v}\alpha}}\left(\frac{\beta}{H_n}\right)^{-1}\right]\,.
\end{aligned}\end{equation} Here $c_s$ is the speed of sound in the plasma, given by $c_s=\sqrt{1/3}$ in a relativistic fluid, and $\kappa_{\rm v}$ is the ratio of the bulk kinetic energy to the vacuum energy, fitted in Ref.~\cite{Espinosa:2010hh} as function of $\alpha$ and $v_w$. The turbulence contribution can be written as~\cite{Ellis:2020awk} \begin{equation} \begin{aligned}\label{eq: turb GW power spectrum}
\Omega_{\rm turb}(\nu)h^2 \simeq & 3.32\times 10^{-4} v_w (1-H_n\tau_{\rm sw}) \left(\frac{100}{g_*}\right)^{1/3} \left(\frac{\beta}{H_n}\right)^{-1} \left(\frac{\kappa_{\rm v}\alpha}{1+\alpha}\right)^{2/3} \left(\frac{\nu}{\nu_{\rm turb}}\right)^3 \\&\times \left[1+\frac{\nu}{\nu_{\rm turb}}\right]^{-11/3}\left[1+\frac{43.98}{v_w}\left(\frac{\beta}{H_n}\right) \left(\frac{\nu}{\nu_{\rm turb}}\right)\right]^{-1}\,,
\end{aligned}\end{equation} where the peak frequency is given by \begin{equation} \begin{aligned}\label{eq: turb peak frequency}
\nu_{\rm turb}=2.89\times 10^{-5}~\text{Hz}\left(\frac{T_n}{100~\text{GeV}}\right)(1+\alpha)^{1/4}\left(\frac{g_*}{100}\right)^{1/6}\frac{1}{v_w}\left(\frac{\beta}{H_n}\right)\,.
\end{aligned}\end{equation} In the following calculations, the significant contribution source
of the GWs is from the sound waves, while the turbulence contribution is a subleading source for the GWs. 

%from bubble collisions in the envelope approximation (found in Ref.~\cite{Huber:2008hg}), acoustic waves (in Ref.~\cite{Hindmarsh:2015qta}) and Kolmogorov-type turbulence (in Ref.~\cite{Caprini:2009yp,Binetruy:2012ze}). The full expressions of these spectra are given in Ref.~\cite{Chen:2017cyc}. 

\begin{table}[tb]
\begin{center}
{\renewcommand{\arraystretch}{1.4}% for the vertical padding
\begin{tabular}{c|ccc|ccc|cc}
\hline\hline
    &  $ \lambda_\eta $ & $\lambda_{h\eta} $  & $m_\eta  $~[GeV] & $ \alpha $ & $ \beta/H_n $ & $v_w$ & $ T_c $~[GeV] & $ T_n $~[GeV]  \\
    \hline\hline
 $ f=1.0 $~TeV  & $0.17$ & $0.19$&  $64.6$ & $ 0.468 $ & $ 209.3 $  & $0.592$ & $100.3$ &$ 48.7 $  \\ 
\hline
 $ f=2.5 $~TeV  & $ 0.44 $ & $ 0.35 $& $105.2$  & $ 0.236 $ & $ 114.7 $ & $0.603 $ &  $97.1$ &$ 54.7 $ \\
  & $ 0.13 $ & $ 0.18 $& $71.7$  & $ 1.291 $ & $ 194.7 $ & $0.795$  &  $97.8$ &$36.2$ \\
   & $ 0.46 $ & $ 0.50 $& $146.9$  & $ 0.165 $ & $ 139.3 $ & $0.592$  &  $104.5$ &$61.5$ \\
\hline\hline
\end{tabular} }
\end{center}
\caption{The parameters of the parameter points marked with purple stars in Figures~\ref{fig: figure 1},~\ref{fig: figure 2} and~\ref{fig: figure 3} which represent SFOEWPTs resulting in GWs that may be measured by the planned LISA space probe. These parameter points are identified in Figure~\ref{fig: The GW signals} by the GW signal strength curves exceeding the LISA sensitivity (the purple line). } \label{tab: data points}
\end{table}
In Figure~\ref{fig: alpha beta Tc Tn}, the values of the parameters $ \alpha $ and $ \beta/H_n $ (left) and $ T_c $ and $ T_n $ (right panel) are plotted for the points resulting in SFOEWPTs, which are depicted by the stars and points in Figures~\ref{fig: figure 1},~\ref{fig: figure 2} and~\ref{fig: figure 3}. The blue and red points represent the SFOEWPT points with $ f=1.0 $~TeV and $ f=2.5 $~TeV, respectively. From these plots, we observe that $ \beta/H_n $ decreases as $ \alpha $ increases leading to stronger phase transitions, while $ T_c \geq T_n $ as expected. The differences between $ T_c $ and $ T_n $ are largest for approximately the points associated with the smallest values of $ \beta/H_n $, thus resulting in the strongest phase transitions and GW signals.  

Finally, in Figure~\ref{fig: The GW signals}, the GW signal strengths ($ \Omega_{\rm GW} h^2 $ given by Eq.~(\ref{eq: total power spectrum})) as a function of the GW frequency $ \nu $ are shown for the points triggering SFOEWPTs, depicted by the stars and points in Figures~\ref{fig: figure 1},~\ref{fig: figure 2} and~\ref{fig: figure 3}. 
Again, the blue and red curves represent the SFOEWPT points with $ f=1.0 $~TeV (upper) and $ f=2.5 $~TeV (lower panel), respectively. It is clear that the generated GW signals from the CH dynamics are testable for most future detectors. As mentioned in the previous section, the parameter points marked with stars in Figures~\ref{fig: figure 1},~\ref{fig: figure 2} and~\ref{fig: figure 3} represent SFOEWPTs resulting in GWs that may be measured by the planned LISA space probe. These parameter points are identified in Figure~\ref{fig: The GW signals} by the GW signal strength curves exceeding the LISA sensitivity (the purple line). The parameters of these points are listed in Table~\ref{tab: data points}.

%%%%%%%%%%%%%%%%%%%%%%%%%%%%%%%%%%%%%%%%%%%%%%%%%%%%%%%%%%%%%%%%%%%%%%%%%%%
\section{Conclusions}
\label{sec: Conclusions}

In this paper, we have presented a general framework for investigating gravitational waves (GWs) from composite Higgs models with SU(N)/Sp(N) cosets arising from underlying
four-dimensional gauge-fermion theories with fermion mass generation from partial
compositeness and where the vacuum is aligned in the direction of a single PNGB Higgs
candidate. The composite dynamics dynamically generates EWSB and potentially alleviates
the naturalness and triviality problems in the SM Higgs sector. We presented an
effective theory taking into account the composite Higgs potential contributions from composite spin-1 resonances, fermion resonances and vector-like masses of the new strongly interacting fermions. 
The total potential of the composite Higgs candidate, $ h $, and the pseudo-scalar partner state, $ \eta $, generated by the composite dynamics allows exploring the possibility of a strong first-order electroweak phase transition (SFOEWPT) when taking the finite temperature corrections into account. 

By scanning the parameter space, we find a large number of points yielding such a strong phase transition, especially if the non-perturbative coefficient $ C_{LR} $ (defined in Eq.~(\ref{eq: pot from top int UV 2})) is positive. Using these parameter points, we make predictions for the models by taking into account GW signals expected from the SFOEWPT, and the projected sensitivities of upcoming experiments. We show how most SFOEWPT points fall within reach of planned future experiments giving the reason for additional excitement. The GW signals generated during the finite temperature EWSB phase transition depend on the number of the EW-charged hyper-fermions ($N$ Weyl fermions). The content of SM-singlet hyper-fermions is irrelevant. Therefore, the more involved CH models~\cite{Cai:2018tet,Alanne:2018xli,Cai:2019cow,Cai:2020bhd,Cacciapaglia:2021aex,Rosenlyst:2021elz} invoking such SM-singlet hyper-fermions to provide composite DM candidates may also be tested by the GW results in this paper.

%%%%%%%%%%%%%%%%%%%%%%%%%%%%%%%%%%%%%%%%%%%%%%%%%%%%%%%%%%%%%%%%%%%%%%%%%%%

\section*{Acknowledgements}
MTF and MR acknowledge partial funding from The Independent Research Fund Denmark, grant numbers DFF 6108-00623 and DFF 1056-00027B, respectively. KT acknowledges the financial support from Academy of Finland, project $\# 342777$. MET acknowledges funding from Augustinus Fonden, application $\#22-19584$, to cover part of the expenses associated with visiting
the University of Helsinki for half a year. We would also like to
thank Marek Lewicki for useful discussions about the gravitational wave calculations.

%%%%%%%%%%%%%%%%%%%%%%%%%%%%%%%%%%%%%%%%%%%%%%%%%%%%%%%%%%%%%%%%%%%%%%%%%%%%%%%%%%%%%%%%%%%%%%%%%%%%
%
\bibliography{GW_CH.bib}
\bibliographystyle{JHEP}
%
%%%%%%%%%%%%%%%%%%%%%%%%%%%%%%%%%%%%%%%%%%%%%%%%%%%%%%%%%%%%%%%%%%%%%%%%%%%%%%%%%%%%%%%%%%%%%%%%%%%% 

\end{document}